\documentclass[fleqn,usenatbib]{mnras}
\hypersetup{linkcolor=red}          
\usepackage{multirow}
\usepackage[dvipsnames]{xcolor}
\usepackage{newtxtext,newtxmath}
\usepackage{orcidlink}

\def\LCDM{$\Lambda${CDM}\ }
\def\galform{{\sc galform} }

\def\equationautorefname~#1\null{Eq.~#1\null}

\title[GALFORM and JWST]{A comparison of pre-existing $\Lambda$CDM predictions with the abundance of {\it JWST} galaxies at high redshift}

\author[S. Lu et al.]
{Shengdong Lu\orcidlink{0000-0002-6726-9499}\thanks{E-mail: \url{shengdong.lu@durham.ac.uk}},
Carlos S. Frenk\orcidlink{0000-0002-2338-716X},
Sownak Bose\orcidlink{0000-0002-0974-5266},
Cedric G. Lacey\orcidlink{0000-0001-9016-5332},
Shaun Cole\orcidlink{0000-0002-5954-7903}, 
\and
Carlton M. Baugh\orcidlink{0000-0002-9935-9755},
John C. Helly\orcidlink{0000-0002-0647-4755}
\\
\\
Institute for Computational Cosmology, Department of Physics, University of Durham, South Road, Durham, DH1 3LE, UK\\
}
\date{}
\pubyear{2024}

\begin{document}
\label{firstpage}
\pagerange{\pageref{firstpage}--\pageref{lastpage}}
\maketitle
\begin{abstract}
Observations with the {\it James Webb Space Telescope} have revealed a high abundance of bright galaxies at redshift, $z\gtrsim 12$, which has been widely interpreted as conflicting with the $\Lambda$CDM model. In Cowley et al. (2018) predictions were made -- prior to the {\it JWST} observations -- for the expected abundance of these galaxies using the Durham semi-analytic galaxy formation model, {\sc galform}, which is known to produce a realistic population of galaxies at lower redshifts including the present day. Key to this model is the assumption of a ``top-heavy" initial mass function of stars formed in bursts (required to explain the number counts and redshift distribution of sub-millimetre galaxies). Here, we compare the rest-frame ultraviolet luminosity functions derived from {\it JWST} observations with those predicted by the Cowley et al. model up to $z=14$ and make further predictions for $z=16$. We find that below $z\sim 10$, the Cowley et al. predictions agree very well with observations, while agreement at $z\gtrsim12$ requires extending the model to take into account the timescale for the growth of obscuring dust grains at these very early times and its dependence on gas metallicity. We trace the evolution of these galaxies from $z=14$ to $z=0$ and find that their descendants typically reside in halos with a median mass $2.5\times 10^{13}\,h^{-1}\,\mathrm{M_{\odot}}$. The stellar masses of the descendants range from $3.2\times 10^{6}\,h^{-1}\,\mathrm{M_{\odot}}$ to $3.2\times 10^{11}\,h^{-1}\,\mathrm{M_{\odot}}$. Although these galaxies were all central galaxies at $z=14$, over half of their descendants end up as satellites in massive halos.
\end{abstract}


\section{Introduction}
\label{sec:introduction}

The $\Lambda$-cold dark matter ($\Lambda$CDM) cosmological model, proposed in the 1980s \citep{Peebles_82,DEFW_85}, has been singularly successful in accounting for a wide variety of observations over a large range of scales, from the temperature structure of the cosmic microwave background radiation \citep[e.g][]{Planck_20} to the luminosity function of the faint satellites of the Milky Way \citep{Bullock_00,Benson_02,Somerville_02}. In some cases, such as the first of these examples, these were genuine theoretical predictions that preceded the observations.

In the $\Lambda$CDM model, galaxies form within dark matter halos that evolve due to gravitational instability in the hierarchical way dictated by the initial power spectrum of density perturbations \citep{Frenk_85}. Thus, the first galaxies form at high redshift in halos of relatively small mass which are, nevertheless, large enough for atomic hydrogen to cool within them \citep{White_Rees_78,Blumenthal_84, White_et_al.(1991),Benitez-Llambay-Frenk_20}. The {\it James Webb Space Telescope} ({\it JWST}) was designed to search for these early galaxies. Detailed theoretical predictions for what the {\it JWST} was expected to find in deep galaxy surveys according to the \LCDM model were published by \cite{Cowley_18} several years before the telescope was launched. That study presented the expected ultraviolet luminosity functions (UVLFs) in the {\it JWST} bands at high redshift, from $z=7$ to $z=16$, as calculated using the \galform semi-analytic model of galaxy formation \citep{Cole_et_al.(2000),Lacey_et_al.(2016)}. Other predictions followed, using semi-analytic models \citep[e.g.][]{Yung2019} or cosmological hydrodynamical simulations \citep[e.g.][]{Ma2019,Vogelsberger2020}. 

Soon after the launch of {\it JWST}, a spate of papers was published claiming the detection of galaxies at very early times based on photometric redshifts. For example, \cite{Donnan_et_al.(2023a)} claimed to have discovered a galaxy at a record redshift of $z=16.4$ (CEERS-93316), only for the subsequent measurement of a spectrum to show that this galaxy is actually at $z=4.91$ \citep[see also][]{Zavala_23,Harikane_et_al.(2024)}. Similarly, \citet{Harikane_et_al.(2023)} claimed to have found a galaxy at $z=16.41$ (S5-z16-1) which, however, as the authors themselves pointed out, has a spectral line detected by ALMA, placing it at $z=4.61$ \citep{Fujimoto_23}. Nevertheless, a number of the high redshift candidates have been spectroscopically confirmed to have $z>11$ (e.g. \citealt{Arrabal2023,Castellano2024,Hainline2024,Harikane_et_al.(2024),Zavala_et_al.(2024)}). 

Several of these early studies published UVLFs at high redshift, based mostly on photometric redshifts \citep[e.g.][]{Bouwens_et_al.(2023a),Donnan_et_al.(2023a),Finkelstein_et_al.(2023),Harikane_et_al.(2023),Morishita_et_al.(2023),Perez-Gonzalez_et_al.(2023b),Robertson_et_al.(2023),Willott_et_al.(2023)}. Most of these papers claimed that the inferred UVLFs were inconsistent with $\Lambda$CDM. For example, \cite{Finkelstein_et_al.(2023)} say: ``With this sample we confirm at higher confidence early {\it JWST} conclusions that bright galaxies in this epoch are more abundant than predicted by most theoretical models.'' Similarly, \cite{Labbe_et_al.(2023)} state that ``... these stellar mass densities are difficult to realize in a standard \LCDM cosmology ...''. Unfortunately, all these papers overlooked the pre-existing \LCDM predictions.

\citet{Boylan-Kolchin_et_al.(2023)} published a simple calculation based on the mass function of dark matter halos which he compared to the stellar mass function inferred by \citet{Labbe_et_al.(2023)}. Boylan-Kolchin found that the baryons in halos would have turn into stars with an efficiency which has, implausibly, to be close to 1 (compared to $\sim 5\%-10\%$ in the Milky Way, e.g. \citealt{Eke2005}) in order to match the data, leading him to conclude that there is a tension with the \LCDM model. This comparison is not trivial because it is the broadband spectral energy distribution (SED), not the stellar mass, that is observed, and inferring the stellar mass from the SED requires making assumptions that are highly uncertain (e.g. regarding the initial mass function and dust obscuration). A similar conclusion was reached from a similar calculation by \citet{Lovell_et_al.(2023)} who stressed the limitations due to the uncertainties in the stellar masses and redshifts of the {\it JWST} galaxies. Indeed, \cite{Steinhardt_et_al.(2023)} showed that using a new set of templates (including a varying IMF) to fit the SED, the inferred stellar masses of \cite{Labbe_et_al.(2023)} are reduced by a factor of 10-50. \citet{Chworowsky_et_al.(2023)} analyzed {\em JWST} data for a sample of galaxies at $z=4-8$ and, using a similar argument to that of \citet{Boylan-Kolchin_et_al.(2023)}, concluded that while the abundance of galaxies at $z\sim 5$ could be accounted for within $\Lambda$CDM assuming a conventional efficiency of baryon conversion into stars ($\sim 0.14$), the efficiency would again have to become implausibly large at higher $z$ unless their inferred stellar masses had been significantly overestimated.

Two studies have attempted to compare the {\it JWST} high redshift data with physically-based models in the \LCDM cosmology. \citet{Kannan_et_al.(2023)} used the $L=740~{\rm Mpc}$ MillenniumTNG hydrodynamic simulation (scaled to match galaxy properties from smaller-volume, higher-resolution simulations) to study the properties of galaxies at $z>10$, including UVLFs. They found broad agreement with the data up to $z=10$ but for $z=10-12$, they require the model galaxies to be dust-free and, beyond $z=12$, the model, even without dust, underpredicts the observed abundance by about one order of magnitude. They suggested that processes not included in the simulation, such as Population III stars, or a variable stellar initial mass function (IMF), may be required. Another physically-based model is that of \citet{Yung_et_al.(2024)} based on the Santa Cruz semi-analytic model \citep{Somerville_99}. They also predicted galaxy number densities at $z>11$ that are at least one order of magnitude below the observational data, even when they neglect dust obscuration. They also speculated that a top-heavy IMF might ease the discrepancy.

The \galform $\Lambda$CDM-based semi-analytic model of galaxy formation of \citet{Cowley_18} (see also \citealt{Lacey_et_al.(2016),Baugh_et_al.(2019)}) is the result of several years of development. A key innovation was the introduction of a top-heavy IMF for stars that form in bursts \citep{Baugh_05}. This is required to account for the number counts and redshift distribution of sub-millimetre sources which cannot otherwise be easily explained in a model that also accounts for the properties of the local galaxy population. A key feature of this model is the self-consistent calculation of the absorption and emission of radiation by dust which is, of course, required to predict sub-millimetre fluxes (see \citealt{Granato2000}). Most \galform work since 2005 has included these two important features. As we shall see, the top-heavy IMF in bursts and a proper treatment of dust obscuration are essential to understand the UV galaxy luminosity functions measured by {\it JWST}. These bursts generate high luminosities while making only about 30 per cent of the stellar mass locked up in galaxies at the present day in the models analysed in this paper.

The basic philosophy of \galform since it was first presented by \cite{Cole_et_al.(2000)} is to calculate quantities, like luminosities, that can be measured directly from observational data and to fix the inevitable free parameters describing the astrophysical processes of galaxy formation to ensure that the model agrees with a subset of basic observed properties of the galaxy population, mostly at the present day, notably the galaxy luminosity function in various bands and a selection of scaling relations. We then use this fully-specified model to make predictions for observables not included in the calibration, most commonly properties at high redshift. If the model fails, then we update the physics in the model, always ensuring agreement with observations at $z=0$. It was this approach that led us to introduce the top-heavy IMF in bursts 20 years ago. Hydrodynamical simulations subsequently adopted a similar philosophy to model calibration \citep[e.g][]{Vogelsberger_14,Schaye_15} but, of course, it is much easier to update a semi-analytic model than a hydrodynamical simulation (for a recent exploration of the sub-grid parameter space of a hydrodynamical simulation, see \citealt{Kugel2023}). 

Here, we compare the predictions of the \citet{Cowley_18} model with the observed UVLFs from $z=7$ to $z=16$. We derive model scaling relations between fundamental properties such as star formation rate, stellar mass, halo mass, and UV luminosity. We also calculate the fate and properties of the present-day descendants of these early {\it JWST} galaxies.

The paper is organized as follows. In \autoref{sec:method}, we summarize the key features of \galform and the calibration of the model parameters. We compare the model with the observed UVLFs in \autoref{sec:original_results}. In \autoref{sec:update_model}, we discuss the role of dust attenuation (\autoref{sec:results-dust}) and the IMF (\autoref{sec:results-imf}) and compare our results with previous studies (\autoref{sec:results-compare}). In \autoref{sec:descendants}, we trace the high-redshift galaxies to the present and study the properties of their descendants. A discussion of our results and our conclusions are presented in~\autoref{sec:conclusion}.

\section{Method}
\label{sec:method}
\subsection{The \galform semi-analytic model}
\label{sec:galform}
The Durham semi-analytic model of galaxy formation, {\sc galform}, was introduced by \citet{Cole_et_al.(2000)}, building on earlier work by \citet{White_Rees_78}, \citet{White_et_al.(1991)}, and \citet{Cole_et_al.(1994)}. Semi-analytic models calculate physical processes along subhalo merger trees which can either be extracted from dark matter-only (DMO) simulations (e.g. \citealt{Helly_et_al.(2003),Jiang_et_al.(2014)}) or generated by Monte-Carlo methods (e.g. \citealt{Cole_et_al.(2000),Parkinson_et_al.(2008)}). \galform solves sets of coupled differential equations that describe gas cooling in halos, star formation, galaxy mergers, feedback from stars and black holes, as well as chemical enrichment and recycling of metals. It combines the predicted star formation histories of galaxies and their chemical evolution with a stellar population synthesis model to build a composite spectral energy distribution for each galaxy. For further details of how these processes are modelled in the most up-to-date version of {\sc galform}, we refer the reader to \citet{Lacey_et_al.(2016)}.

\subsubsection{The dark matter-only simulations}
\label{sec:dmo}
\citet{Cowley_18} applied \galform to the P--Millennium simulation, a variant of the Millennium simulation \citep{Springel_et_al.(2005)}, with cosmological parameter values derived from \emph{Planck} data  \citep{Planck2016}, in a cube of sidelength $800\,\mathrm{Mpc}$. The P--Millennium simulation was introduced by \cite{Baugh_et_al.(2019)}. Halos were initially identifed with a friends-of-friends algorithm \citep{DEFW_85} and self-bound structures were then identified with the {\sc subfind} algorithm \citep{Springel_et_al.(2001)}. Merger trees were constructed from these subhalo catalogues using the `DHalos' algorithm \citep{Jiang_et_al.(2014)}.

The large volume of P--Millennium allows us to study the rare massive halos that could host bright galaxies at high redshift, $z\sim 14-16$, but the relatively large dark matter particle mass, $1.06\times 10^{8}\,h^{-1}\,\mathrm{M_{\odot}}$, limits the ability to follow the low-mass halos that could host very faint galaxies. To overcome this resolution limitation,  we combine the P--Millennium with the EAGLE--DMO (dark matter-only) simulation \citep{Guo_et_al.(2016)}. EAGLE--DMO  assumed the same cosmological parameters as the P--Millennium, but it has much higher mass resolution, a particle mass of $1.15 \times 10^{7}\,h^{-1}\,\mathrm{M_{\odot}}$, 10 times smaller than P--Millennium. To combine the two simulations, galaxies with halo masses larger than $5\times 10^{9}\,h^{-1}\,\mathrm{M_{\odot}}$ (roughly corresponding to 50 DM particles in P--Millennium and 450 DM particles in EAGLE--DMO) are taken from P--Millennium, while those with lower halo masses are taken from the EAGLE--DMO simulation. 

Since the EAGLE--DMO simulation has a cube sidelength of $100 \, {\rm Mpc}$, $1/8$ that of P--Millennium, when calculating the luminosity functions we assign a weight $8^3=512$ larger to the EAGLE--DMO galaxies than to the P--Millennium galaxies. We will return to this point in \autoref{sec:original_results}. With this combination, we are able not only to study high redshift rare massive objects, but also low-mass halos down to $\sim 10^{9}\,h^{-1}\,\mathrm{M_{\odot}}$ (about 100 DM particles in EAGLE--DMO).

\begin{figure}
\centering
\includegraphics[width=\columnwidth]{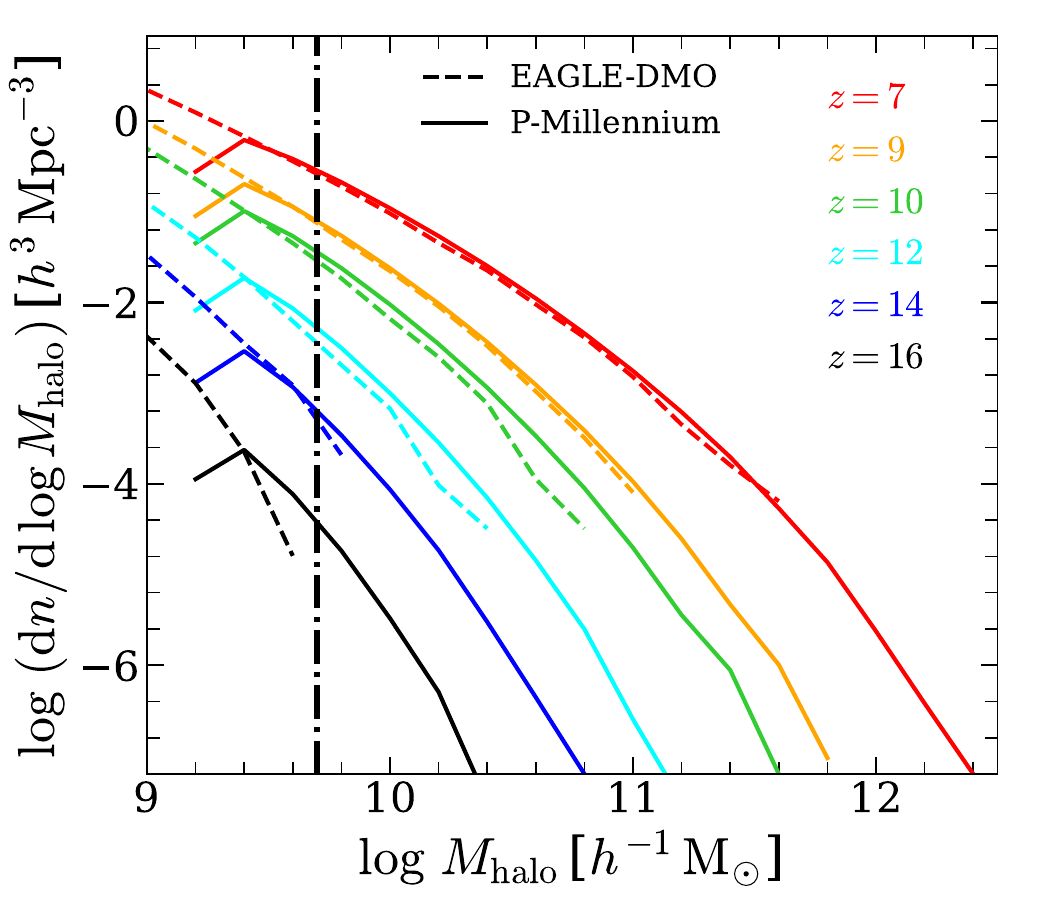}\vspace{-0.5cm}
\caption{Halo mass functions measured in the P--Millennium (solid) and EAGLE--DMO (dashed) N-body simulations  between $z=7$ and $z=16$. The vertical dashed-dotted line indicates $M_{\rm halo}=5\times 10^{9}\,h^{-1}\,\mathrm{M_{\odot}}$, which corresponds to the mass of 50 particles in P--Millennium.}
\label{fig:hmf}
\end{figure}

For P--Millennium, there are 271 snapshots in total, 15 between $z=12$ and $z=16$; for EAGLE--DMO, there are 400 snapshots in total, 14 between $z=12$ and $z=16$. This allows us accurately to capture the merging history of halos and galaxies. In \autoref{fig:hmf}, we show the halo mass functions in both P--Millennium (solid) and EAGLE--DMO (dashed) from $z=7$ to $z=16$. We observe that even at $z=16$, halos as massive as $2\times 10^{10}\,h^{-1}\,\mathrm{M_{\odot}}$ have a number density as high as $\sim 10^{-7}\,h^{3}\,\mathrm{Mpc^{-3}}$, which corresponds to 16 halos/dex in the simulation cube.

Throughout we assume a flat universe and, to be consistent with the P--Millennium and EAGLE--DMO, we adopt cosmological parameters from \citet[][i.e. $\Omega_{\rm m}=0.307$, $\Omega_{\Lambda}=0.693$, $\Omega_{\rm b}=0.0483$, $\sigma_{8}=0.8288$, and $h=0.6777$]{Planck2016}.

\subsubsection{Supernova feedback model}
\label{sec:sn_model}
Energy injected by supernovae heats up galactic gas, altering its ability to cool and producing a feedback loop that limits further star formation. This is one of the key processes that determines galaxy evolution. In the fiducial \galform model (e.g. \citealt{Lacey_et_al.(2016)}), this process is described by:
\begin{equation}
\label{eq:sn_eject}
\dot{M}_{\mathrm{eject}}=\beta\left(V_{\mathrm{c}}\right) \psi=\left(V_{\mathrm{c}} / V_{\mathrm{SN}}\right)^{-\gamma_{\mathrm{SN}}} \psi,
\end{equation}
where $\dot{M}_{\mathrm{eject}}$ is the cold gas ejection rate; $\beta$ is the ``mass loading'' factor determined by the circular velocity, $V_{\mathrm{c}}$, of the disc or bulge, depending on whether stars are forming quiescently or in a starburst; $\psi$ is the star formation rate (SFR); $V_{\mathrm{SN}}$ and $\gamma_{\mathrm{SN}}$ are adjustable parameters which are set here to $V_{\mathrm{SN}}=320\,\mathrm{km\,s^{-1}}$ and $\gamma_{\mathrm{SN}}=3.4$ in order to match selected properties of the local galaxy population (see \autoref{sec:model_design} for details of the model calibration). The ejected gas accumulates in a reservoir of mass $M_{\rm res}$ outside the halo, and gradually rejoins a hot gas corona within the halo virial radius at a rate:
\begin{equation}
\label{eq:sn_return}
\dot{M}_{\text {return }}=\alpha_{\text {ret }} \frac{M_{\text {res }}}{\tau_{\text {dyn,halo }}},
\end{equation}
where $\tau_{\text {dyn,halo }}$ is the halo dynamical time and $\alpha_{\text {ret }}$ is a constant, set to $\alpha_{\text {ret }}=1.0$. We refer to this as the standard (STD) feedback model.

A deficiency of the STD model is that it does not produce enough ionizing photons at early times to reionize the Universe at the high redshift implied by the \citet{Planck2016} data. This could be remedied by reducing the strength of supernova feedback at high redshift, leading to enhanced star formation at early times. One consequence of this is that, with the original parametrization of feedback, this would result in the overproduction of dwarf galaxies today as well as in an incorrect stellar mass-metallicity relation for these galaxies. This led \citet{Hou_et_al.(2016)} to propose a new SN feedback model for \galform in which the feedback strength changes with cosmic time. In this model, referred to as EvolFB, the simple power-law ``mass loading'' factor, $\beta (V_{\rm c})$ (\autoref{eq:sn_eject}) is replaced by a broken power law as a function of $V_{\rm c}$:
\begin{equation}
\label{eq:sn_updated_beta}
\beta=\left\{\begin{array}{l}
\left(V_{\mathrm{c}} / V_{\mathrm{SN}}\right)^{-\gamma_{\mathrm{SN}}} \,\,V_{\mathrm{c}} \geq 50\,\mathrm{km\,s^{-1}} \\
\left(V_{\mathrm{c}} / V_{\mathrm{SN}}^{\prime}\right)^{-\gamma_{\mathrm{SN}}^{\prime}}\,\, V_{\mathrm{c}}<50\,\mathrm{km\,s^{-1}}
\end{array} ,\right.
\end{equation}
where $V_{\mathrm{SN}}$ depends on redshift as:
\begin{equation}
\label{eq:sn_updated_vsn}
V_{\mathrm{SN}}\left(\mathrm{km}\,\mathrm{s}^{-1}\right)=\left\{\begin{array}{cc}
180 & z>8 \\
-35 z+460 & 4 \leq z \leq 8 \\
320 & z<4
\end{array} .\right.
\end{equation}
The index $\gamma^{\prime}_{\mathrm{SN}}$ is taken to be $\gamma^{\prime}_{\mathrm{SN}}=1.0$ and $V^{\prime}_{\mathrm{SN}}$ is fixed by the condition that the two power laws should join at $V_{\rm c}=50\,\mathrm{km\,s^{-1}}$. In this new model the strength of SN feedback is weaker at $z>4$ than in the original model allowing \galform simultaneously to match: (1) the redshift, $z_{\rm re, half}$, at which the Universe was 50 per cent reionized according to \cite{Planck2016}, (2) the luminosity function of Milky Way satellites and (3) the stellar mass-metallicity relation for Milky Way satellites. 

\subsubsection{Stellar initial mass function}
\label{sec:imf_model}
The stellar initial mass function (IMF) describes the mass distribution of stars at birth. Its form affects the evolution of gas, stars, and metal content of galaxies. The IMF is typically described as a power-law: 
\begin{equation}
\label{eq:imf}
\Phi(m)=\frac{\mathrm{d} N}{\mathrm{~d} \ln m} \propto m^{-x},
\end{equation}
where $N$ is the number of stars as a function of their mass at birth, $m$. In \galform we distinguish between two modes of star formation (SF): (i) a quiescent (or disc) SF with a \citet{Kennicutt(1983)} IMF ($x=0.4$ for $m<1\,\mathrm{M_{\odot}}$ and $x=1.5$ for $m>1\,\mathrm{M_{\odot}}$), and (ii) a starburst mode with a top-heavy IMF ($x=1$ for the entire mass range, $0.1\,\mathrm{M_{\odot}}$ to $100\,\mathrm{M_{\odot}}$). As we mentioned in the Introduction (\autoref{sec:introduction}), \cite{Baugh_05} found that a top-heavy IMF in starbursts is necessary to account for the number counts and redshift distribution of sub-millimetre galaxies (SMGs), while remaining consistent with observational constraints at $z=0$ (see also \citealt{Lacey_et_al.(2016)}). 

Since the frequency of starbursts is higher at higher redshift, the dual IMF in \galform is, in practice, equivalent to a redshift-dependent IMF which, on average, becomes increasingly top-heavy at higher redshift. We note that although a redshift dependence of the IMF remains an open question, evidence has accumulated suggesting the presence of a top-heavy IMF in the early Universe. For example, \citet{Sneppen_et_al.(2022)} fitted a temperature-dependent IMF to galaxies in the COSMOS2015 catalogue \citep{Laigle_et_al.(2016)} and found that the IMF becomes progressively top-heavy with increasing redshift. \citet{Cameron_et_al.(2023)} also reported two Lyman-$\alpha$ emitting galaxies at redshifts 5.9 and 7.9 that show evidence for exceptionally top-heavy IMFs. \citet{Liang_et_al.(2021)} found that the IMF tends to be more top-heavy at lower metallicities, supporting the possibility of a top-heavy IMF at high redshift.

\subsubsection{Stellar population and luminosity}
\label{sec:sp}
The stellar population synthesis model in \galform is explained in \citet{Lacey_et_al.(2016)}. Here, we provide a brief overview of the procedure for deriving the properties of the stellar populations and the luminosities in different bands for galaxies in {\sc galform}.    

To obtain galaxy luminosities, \galform calculates spectral energy distributions (SEDs) from a set of single stellar population (SSP) templates. The SED of a stellar population at a given time, $t$, can be written as: 
\begin{equation}
L_{\lambda}(t) = \int_0^t dt^{\prime} \, \int_0^{\infty} dZ^{\prime} \, \Psi(t^{\prime},Z^{\prime}) \,
L^{\rm (SSP)}_{\lambda}(t-t^{\prime},Z^{\prime};\Phi),
\label{eq:L_tot}
\end{equation}
where $\Psi(t^{\prime},Z^{\prime})\, dt^{\prime}\,dZ^{\prime}$ represents the mass of stars at birth which formed in the time interval $(t^{\prime},t^{\prime}+dt^{\prime})$ and metallicity range $(Z^{\prime},Z^{\prime}+dZ^{\prime})$. $L^{\rm (SSP)}_{\lambda}(t,Z;\Phi)$ is the SED of an SSP of one solar  mass with age $t$ and metallicity $Z$, given a specific IMF, $\Phi(m)$. $\Psi(t,Z)$ is derived by summing over the star formation histories of all the progenitor galaxies which merged into the final galaxy. The SSP luminosity is determined by integrating the luminosity, $L^{\rm (star)}(t,Z,m)$, of a star of initial mass $m$, metallicity $Z$, and age $t$ over the IMF:
\begin{equation}
L^{\rm (SSP)}_{\lambda}(t,Z;\Phi) = \int_{m_L}^{m_U} \, L^{\rm (star)}_{\lambda}(t,Z,m) \, \Phi(m) \, d\ln m ,
\end{equation}
where $m_{\rm L}$ and $m_{\rm U}$ are the lower and upper mass limit of the IMF (set to be $0.1\,\mathrm{M_{\odot}}$ and $100\,\mathrm{M_{\odot}}$, respectively; see \autoref{sec:imf_model}). Here, as in the original models, we adopt two different IMFs, one for quiescent (or disc) star formation and another for starbursts (see \autoref{sec:imf_model}). Thus, we apply \autoref{eq:L_tot} separately to the stars formed in the disc and starburst modes, and then add these to get the total luminosities in each galaxy. The SSP library used here is that from \citet{Maraston(2005)}, which has a fine grid of ages, but only a coarse grid of metallicities: $Z=0.001,\,0.01,\,0.02,$ and $0.04$. Thus, we first interpolate $L^{\rm (SSP)}_{\lambda}(t,Z;\Phi)$ in both $t$ and $Z$ as needed and then multiply it by the suitably normalized filter response function and integrate it to obtain the broad-band luminosities and magnitudes of galaxies. Absolute magnitudes are calculated with zero points in either the Vega or AB systems, as required.

\subsubsection{Dust model}
\label{sec:dust_model}
Dust attenuates the luminosities of galaxies and is an important consideration when predicting the galaxy luminosity function in different wavebands. \galform includes a self-consistent dust model that calculates the absorption of radiation by dust in UV, optical, and near-IR bands, as well as the re-radiation of this energy at IR and sub-mm wavelengths. The model assumes a two-phase dust medium, with dust in molecular clouds embedded in a diffuse dust component. The absorption and re-radiation of energy by the two dust components are calculated separately and, thus, the temperature of the different dust components can vary. Some simplifying assumptions are made: (i)~a single dust temperature is assumed for each dust component (cloud or diffuse); (ii)~temperature fluctuations for small grains are ignored; (iii)~the dust opacity is assumed to be a power law at IR/sub-mm wavelengths, and (iv) polycyclic aromatic hydrocarbon (PAH) features are ignored. We refer readers to \citet{Lacey_et_al.(2016)} for a more detailed description of the dust absorption and emission model. 

\citet{Cowley_18} adopted a different approach, using the spectrophotometric radiative transfer code, {\sc grasil} \citep{Silva_et_al.(1998)}, to calculate the dust absorption and emission processes; this is slightly different from the default \galform dust model used in this paper. We confirm, however, that the difference in dust models has a negligible effect on our results (see also \citealt{Cowley2017}).

\subsection{Model parameters}
\label{sec:model_design}

\begin{figure*}
\centering
\includegraphics[width=2\columnwidth]{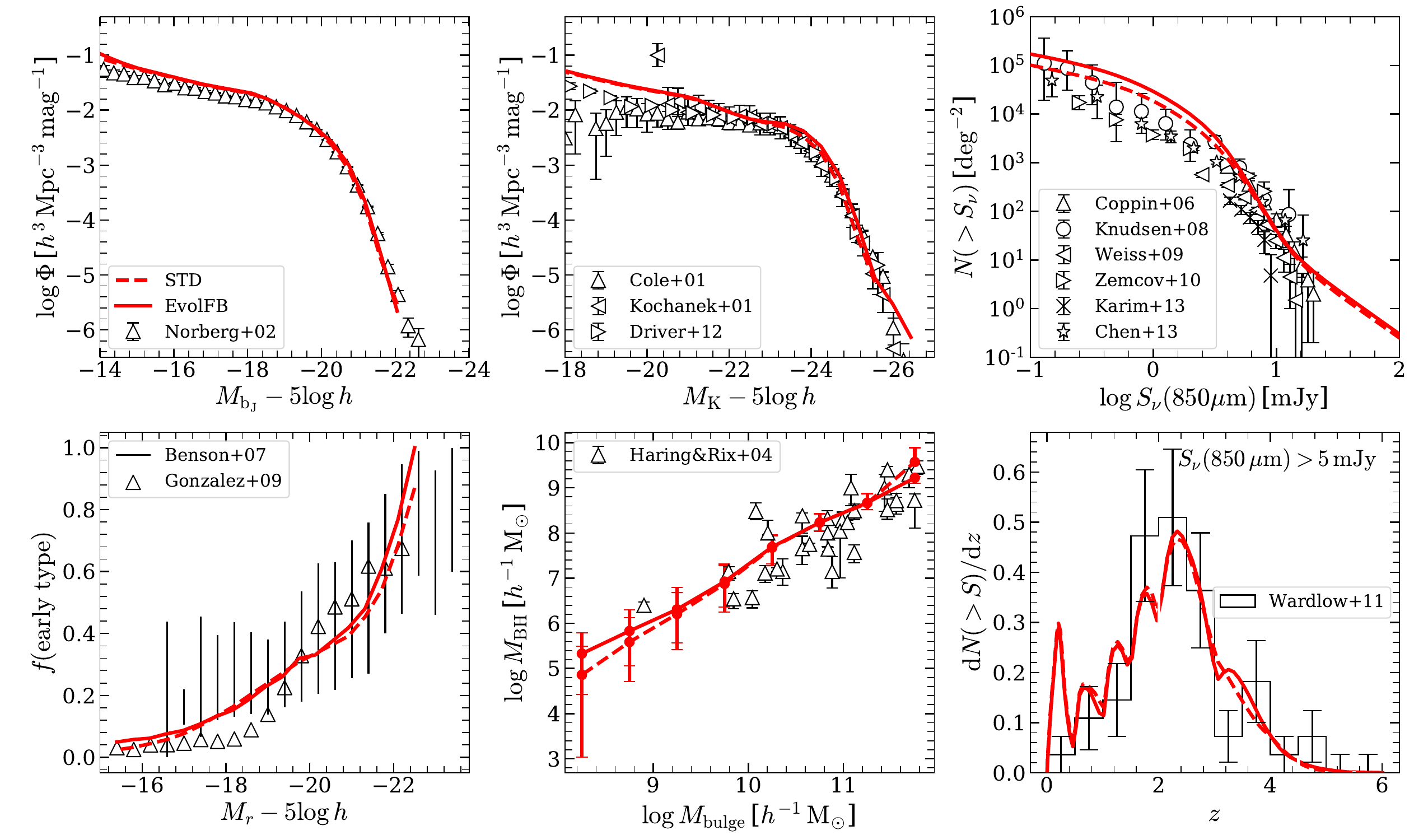}
\caption{Comparison of \galform results with the properties of the local galaxy population used to calibrate the model parameters. Upper left: $b_J-$band Vega luminosity function (LF) at $z=0$ compared to data from \citet{Norberg_et_al.(2002)}; upper middle: $K-$band Vega LF at $z=0$ compared to data from \citet{Cole_et_al.(2001),Kochanek_et_al.(2001),Driver_et_al.(2012)}; upper right: cumulative number counts of sub-millimetre galaxies compared to data from \citet{Coppin_et_al.(2006),Knudsen_et_al.(2008),Weiss_et_al.(2009),Zemcov_et_al.(2010),Karim_et_al.(2013),Chen_et_al.(2013)}; lower left: the morphological fractions of galaxies as a function of their $r-$band luminosity at $z=0$, compared to data from \citet{Benson_et_al.(2007)} (vertical hatched region; based on disc-bulge decomposition) and \citet{Gonzalez_et_al.(2009)} (open squares; based on the Petrosian concentration index in the $r-$band); lower middle: the black hole–bulge mass relation of galaxies at $z=0$, compared to data from \citet{Haring_et_al.(2004)} with the error bars indicating the range from 16th to 84th percentiles (i.e. $1\sigma$ range); lower right: the redshift distribution of sub-millimetre galaxies ($S_{\nu}(850\,\mathrm{\mu m})>5\,\mathrm{mJy}$), compared to data from \citet{Wardlow_et_al.(2011)}. In each panel, the STD model and EvolFB model (\autoref{sec:model_design}) are indicated by red dashed and red solid curves, respectively.}
\label{fig:low_z}
\end{figure*}

\begin{figure*}
\centering
\includegraphics[width=0.9\textwidth]{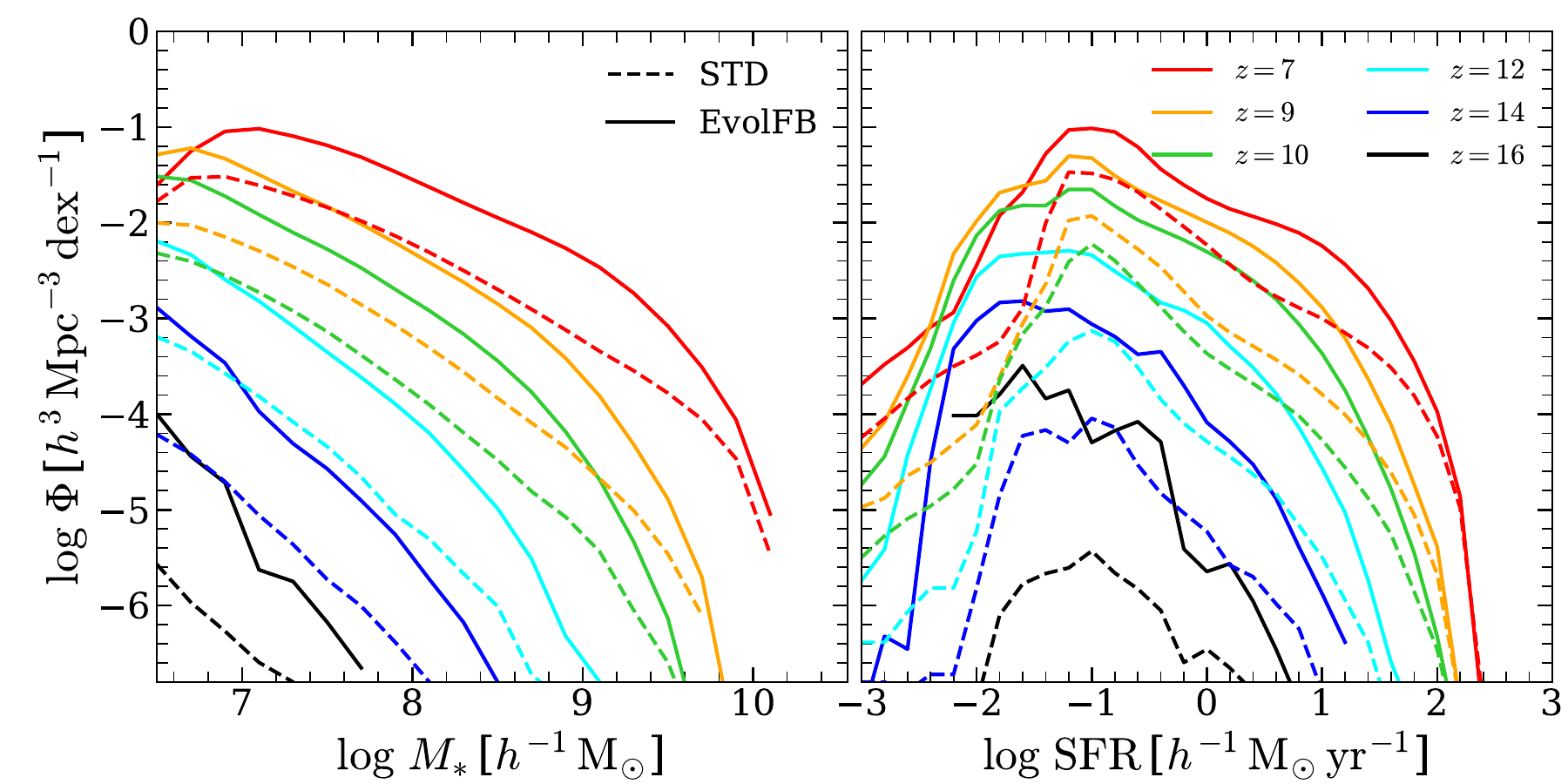}\vspace{-0.1cm}
\caption{\galform predictions for the stellar mass function (left) and the star formation rate (SFR) distribution function (right). In each panel, the results from $z=7$ to $z=16$ are shown using different colours, as indicated in the key. The STD model and EvolFB model are shown with dashed and solid curves respectively.}
\label{fig:mstar_muv_distribution}
\end{figure*}

The \galform model is defined by a set of parameters that describe the various physical processes thought to be relevant to galaxy formation. While this is often a source of criticism of semi-analytic models (although not of hydrodynamical simulations which, generally, have more adjustable parameters), we stress that, contrary to a common misconception, we do not have unlimited freedom in the choice of the parameter values. The equations describing the model are constructed from physical arguments and the approximate range of values allowed for any given parameter is constrained by theoretical considerations or guided by observational constraints. 

\begin{figure*}
\centering
\includegraphics[width=0.9\textwidth]{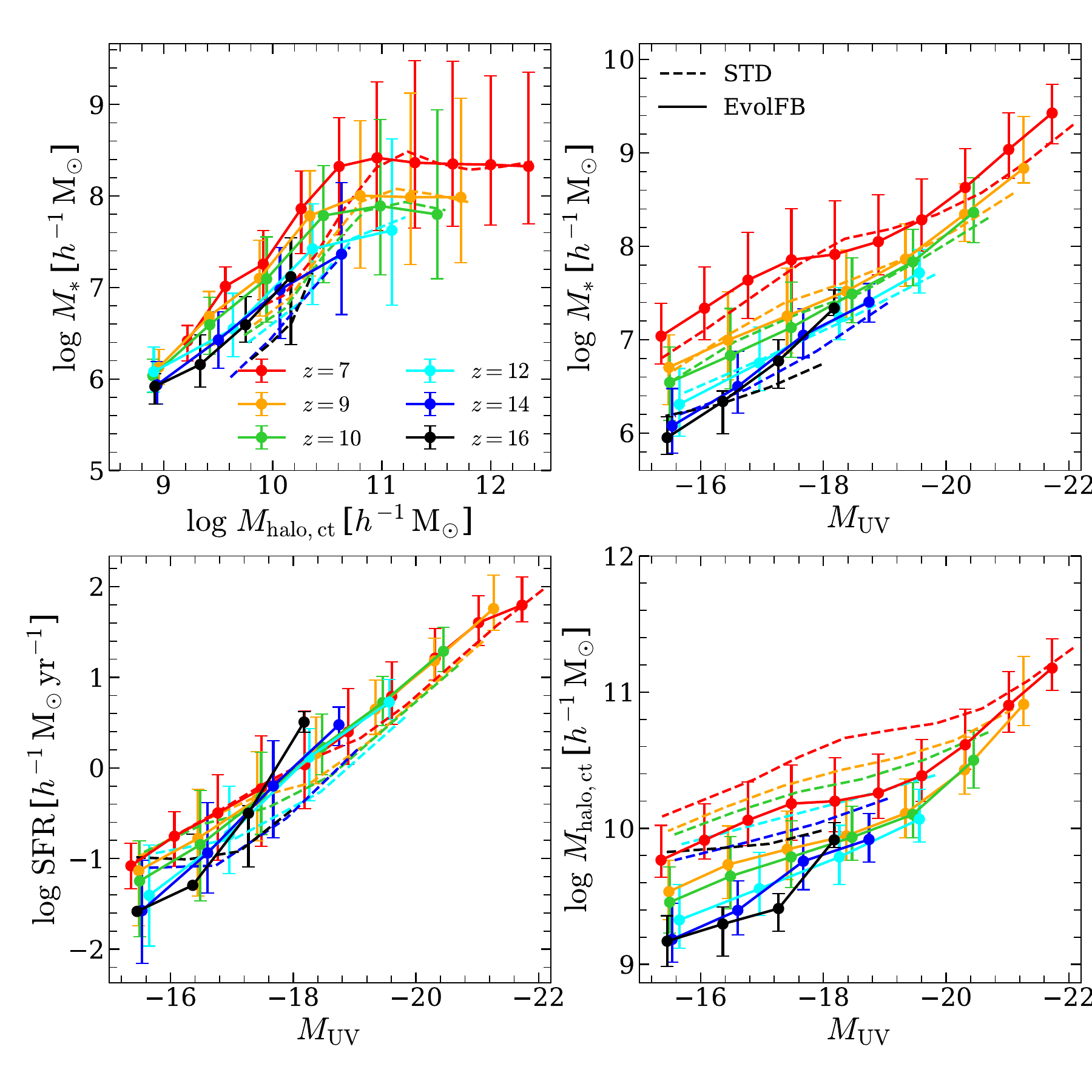}\vspace{-0.2cm}
\caption{\galform predictions. Upper left: the correlation between stellar mass and halo mass; upper right: the correlation between stellar mass and UV magnitude ($M_{\rm UV}\equiv M_{\rm AB}(1500\Angstrom)-5\log\,h$); lower left: the correlation between star-formation rate and UV magnitude; lower right: the correlation between halo mass and UV magnitude. In all panels, only galaxies with UV magnitude $M_{\rm UV}<-15$ are included. The halo mass ($M_{\rm halo,ct}$) used in this figure is the mass of the host halo of the galaxy at  the last snapshot at which it was still a central galaxy. The results from $z=7$ to $z=16$ are shown using different colours, as indicated by the key. The STD model and EvolFB model, respectively, are shown with dashed and solid curves. The error bars indicate the 16th to 84th percentile ($\pm 1\sigma$) range. As both models exhibit a similar level of scatter, for clarity, we only show the error bars for the EvolFB model in the bottom two panels.}
\label{fig:mstar_muv_mhalo}
\end{figure*}

\begin{figure}
\centering
\includegraphics[width=\columnwidth]{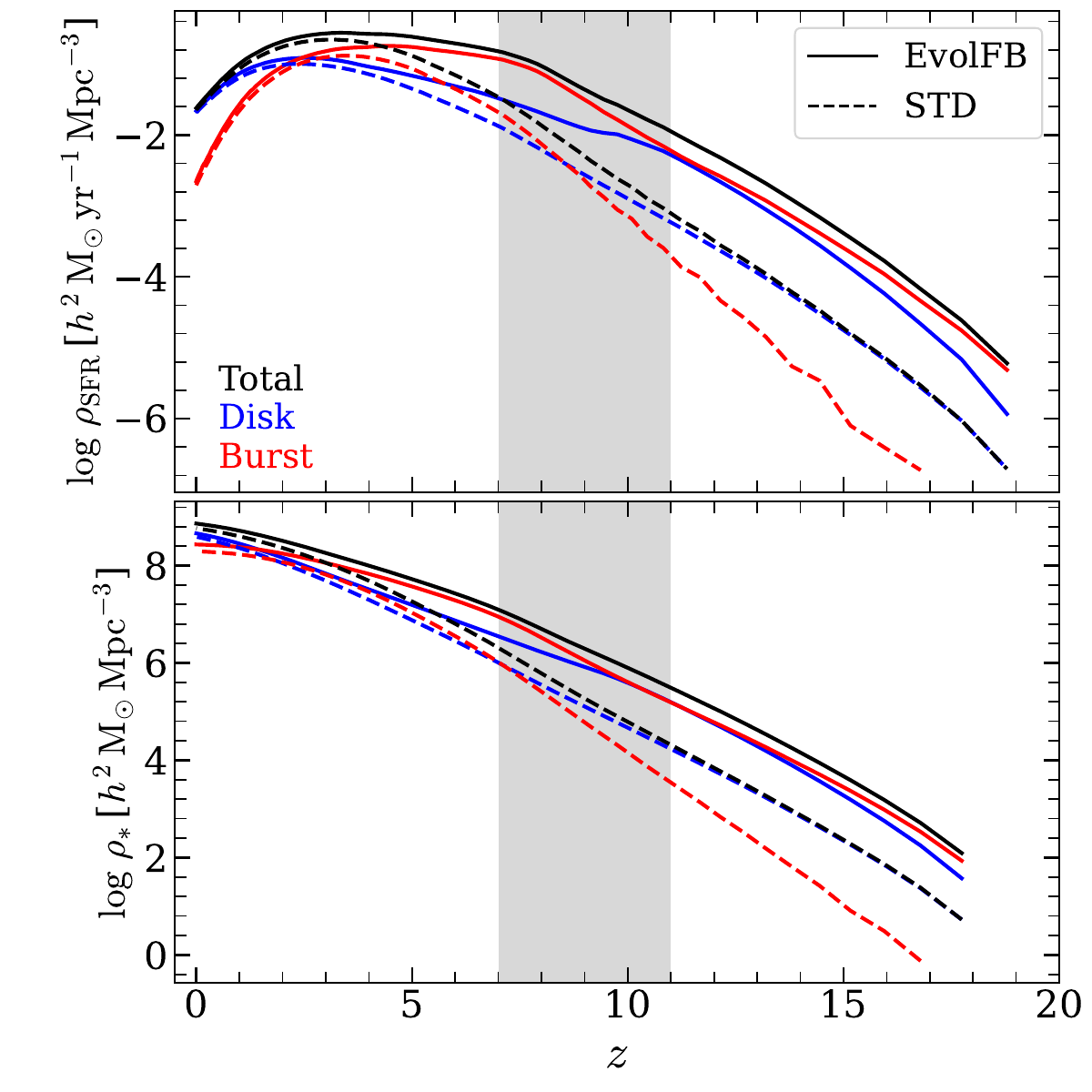}\vspace{-0.4cm}
\caption{Upper panel: the redshift evolution of the star formation rate density (SFRD) from $z=19$ down to $z=0$. The results of the EvolFB model and STD model are indicated by solid and dashed curves respectively, with colours representing the SFRD in different star formation (SF) modes (black for the total SFRD, blue for the SFRD in disc/quiescent SF mode, red for the SFRD in starbursts). Lower panel: the integrated stellar mass density at different redshifts, showing both the EvolFB and STD models, along with the separate contributions from disc and starburst SF, as well as the total. Below $z=7$, the SFRD is measured from the P--Millennium simulation and above $z=11$ from the EAGLE-DMO simulation. In between these redshifts, we apply a weight that changes linearly from 1 to 0 (or 0 to 1) between $z=7$ and $z=11$ for P--Millennium (or EAGLE--DMO), allowing the two simulations to join smoothly.}
\label{fig:sfh}
\end{figure}

Following the philosophy behind {\sc galform}, as set out in \citet{Cole_et_al.(2000)}, \citet{Lacey_et_al.(2016)}, and other \galform papers, we calibrate the model parameters to reproduce a small set of observational constraints, mainly at {\it low} redshift, including: (i)~the $b_J-$ and $K-$band luminosity functions at $z=0$; (ii)~the morphological fractions as a function of $r-$band luminosity of galaxies at $z=0$; (iii)~the black hole–bulge mass relation at $z=0$; (iv)~the number counts and redshift distribution of sub-millimetre galaxies; and (v)~UV galaxy luminosity functions at $z=3$ and $z=6$.

 Here, we calculate the same models as \citet{Cowley_18}. We adopt the same two SN feedback models, STD and EvolFB, described in \autoref{sec:sn_model}, retaining the same values of all the model parameters. For example, we did not change the parameters of the EvolFB model to reflect a recent change in the inferred redshift of reionization from Planck data, from $z_{\rm re}\gtrsim 8.5$ (\citealt{Planck2016}) to $z_{\rm re}\sim 7-8$ (\citealt{Planck_20}).

The main properties of the local galaxy population and of the submillimetre galaxies used to calibrate the model parameters for both the STD and EvolFB models applied to the P--Millennium simulation are shown in \autoref{fig:low_z}. All other properties of the resulting galaxy population are predictions of the model. In particular, the UV luminosity function of galaxies at the redshifts relevant here, $z=7-16$, was {\it not} used to calibrate the model. As we can see from \autoref{fig:low_z}, both the STD model and the EvolFB model show remarkable agreement with the observations at lower redshifts, which is consistent with the results of \citet{Lacey_et_al.(2016)} and \citet{Cowley_et_al.(2019)} for the STD model and of \citet{Hou_et_al.(2016)} for the EvolFB model. The STD model is quite close to the EvolFB model at $z=0$ as the evolving SN feedback converges to the standard case below $z=4$ \citep{Hou_et_al.(2016)}.

\section{{\sc galform} predictions vs observations}
\label{sec:original_results}

\subsection{Galaxy properties}
\label{sec:correlations}
Before comparing \galform predictions with the observational data from {\it JWST}, in \autoref{fig:mstar_muv_distribution} we present the predictions of the \galform model for (i)~the stellar mass function (SMF) and (ii)~the star formation rate (SFR) distribution function. Here, we also combine the P--Millennium and EAGLE--DMO simulations using the method described in \autoref{sec:dmo} in order to mitigate the effects of the low mass resolution of P--Millennium. We can see from the left panel of \autoref{fig:mstar_muv_distribution} that the stellar mass function evolves significantly from $z=7$ to $z=16$. The SMF at the lower redshifts shows both a higher number density at fixed stellar mass and extends to larger masses. Across the whole redshift and stellar mass ranges, the EvolFB model predicts a higher galaxy abundance than the STD model. This is a natural consequence of the redshift-dependent supernova feedback model, in which the feedback becomes weaker at higher redshifts ($z>4$), allowing star formation to continue more efficiently. 

The SFR distribution function (right panel) also shows significant evolution, with the SFR increasing, on average, towards lower redshifts. The SFR distribution function in the EvolFB model has a similar shape to that in the STD model at $z\lesssim 10$. The highest SFRs at $z\gtrsim 12$ hardly exceed $40\,h^{-1}\,\mathrm{M_{\odot}\, yr^{-1}}$ in either model. The SFR distributions here are roughly consistent with those of \citet[][Fig.~3]{Cowley_18}, but we do see slight differences, especially at high redshift  ($z\sim 14-16$). There are more galaxies with low SFR ($<0.1\,h^{-1}\,\mathrm{M_{\odot}\,yr^{-1}}$) in our work. This is because, with the combination of the P--Millennium and EAGLE--DMO simulations, we are able to resolve galaxies with lower masses at high redshifts.

\autoref{fig:mstar_muv_mhalo} shows the correlations between stellar mass, halo mass, star formation rate, and UV magnitude from $z=7$ to $16$. We define:
\begin{equation}
M_{\rm UV} = M_{\rm AB}(1500\Angstrom)-5\log h,
\end{equation}
where $M_{\rm AB}(1500\Angstrom)$ is the absolute AB magnitude of galaxies at $\lambda = 1500\,\Angstrom$ in their rest-frame. When comparing the predictions of the model to observations in the following sections, the observational data are also scaled to the same units. All galaxies, central and satellites, are included in \autoref{fig:mstar_muv_mhalo}. In this figure, for both types of galaxy,  we take as the ``halo mass'' the mass of the halo at the last output time when the galaxy was a central galaxy,  $M_{\rm halo,ct}$. For centrals, this is just the current halo mass; for satellites it is the mass just before infall into the larger host halo. This definition is used only in this figure. Elsewhere in the paper, the halo mass is defined as the mass of the halo in which the galaxy resides, which we denote $M_{\rm halo}$. For centrals this is the same as $M_{\rm halo,ct}$ but for satellites, it is the mass of the current host halo.

As can be seen, both the STD and EvolFB models exhibit an approximate power-law relationship between halo mass and stellar mass below $M_{\rm halo,ct}\sim 10^{11}\,\mathrm{M_{\odot}}$. At higher halo masses, the relation flattens and the scatter increases significantly. At a given halo mass, lower redshift galaxies are more likely to have higher stellar masses. The EvolFB model predicts higher stellar masses than the STD model at $M_{\rm halo,ct}\lesssim 10^{11}\,\mathrm{M_{\odot}}$, but similar stellar masses to the STD model at higher halo masses. For the sample used in this figure, $M_{\rm UV} < -15$, nearly all galaxies are centrals, so the figure would be very similar had we plotted $M_{\rm halo}$ rather than $M_{\rm halo,ct}$.

The correlations between $M_{\rm UV}$ and other properties exhibit a strong redshift dependence in both models. Stellar mass, star formation rate, and halo mass all show a positive correlation with the UV luminosity of the galaxy, as expected. At fixed stellar mass and halo mass, high-redshift galaxies appear to be brighter in the UV than low-redshift galaxies. This shows that the ``bright'' galaxies observed at high redshifts are {\em not} ``massive'' galaxies, contrary to what is often stated in the observational literature \citep[e.g.,][]{Harikane_et_al.(2023),Labbe_et_al.(2023)}. Interestingly, in the EvolFB model, high-redshift galaxies show a steeper SFR-$M_{\rm UV}$ relation than low-redshift galaxies (see the bottom left panel of \autoref{fig:mstar_muv_mhalo}), while in the STD model, this feature is not as obvious.

\autoref{fig:sfh} shows the redshift evolution of the star formation rate density (SFRD; upper panel) and the integrated stellar mass density (lower panel). The integrated stellar mass density is calculated by integrating the SFRD from $z = 19$ to a given redshift and then multiplying by a factor $(1 - R)$ to obtain the mass in long-lived stars and stellar remnants. Here, $R$ represents the recycled fraction corresponding to the assumed IMF: $R = 0.44$ for the \citet{Kennicutt(1983)} IMF in disc quiescent SF mode and $R = 0.54$ for the $x = 1$ IMF in starbursts. The total stellar mass density is the sum of the stellar mass densities formed in quiescent and starburst SF modes. 

As we have seen, the P--Millennium simulation does not resolve low-mass galaxies at high redshift and thus underestimates the SFRD at those redshifts. In this figure, we show the results of P--Millennium below $z=7$ and EAGLE--DMO above $z=11$. In between, we apply a weight that linearly varies from 1 to 0 (or 0 to 1) between $z=7$ and $z=11$ for P--Millennium (or EAGLE--DMO), enabling the two simulations to join smoothly. As shown, the SFRDs of both the EvolFB model and the STD model peak at $z \sim 2 - 4$, with the EvolFB model (solid curve) consistently showing a higher SFRD than the STD model at $z \gtrsim 3$. At $z \lesssim 2.5$, the star formation rate in quiescent mode (blue curves) is significantly higher than that in starbursts (red curves) in both models. However, at higher redshifts ($z \gtrsim 2.5$), the EvolFB model exhibits a significantly higher SFRD in starbursts (red solid) compared to discs (blue solid), while the STD model shows only a slight excess in SFRD in starbursts (red dashed) at $2.5 \lesssim z \lesssim 9$. This figure confirms that the SFRD continues to decrease towards higher redshifts starting from $z \sim 2 - 4$ and does not reach zero, at least not before $z \sim 19$.

\subsection{UV luminosity functions}
\label{sec:uvlfs}

\begin{figure*}
\centering
\includegraphics[width=\textwidth]{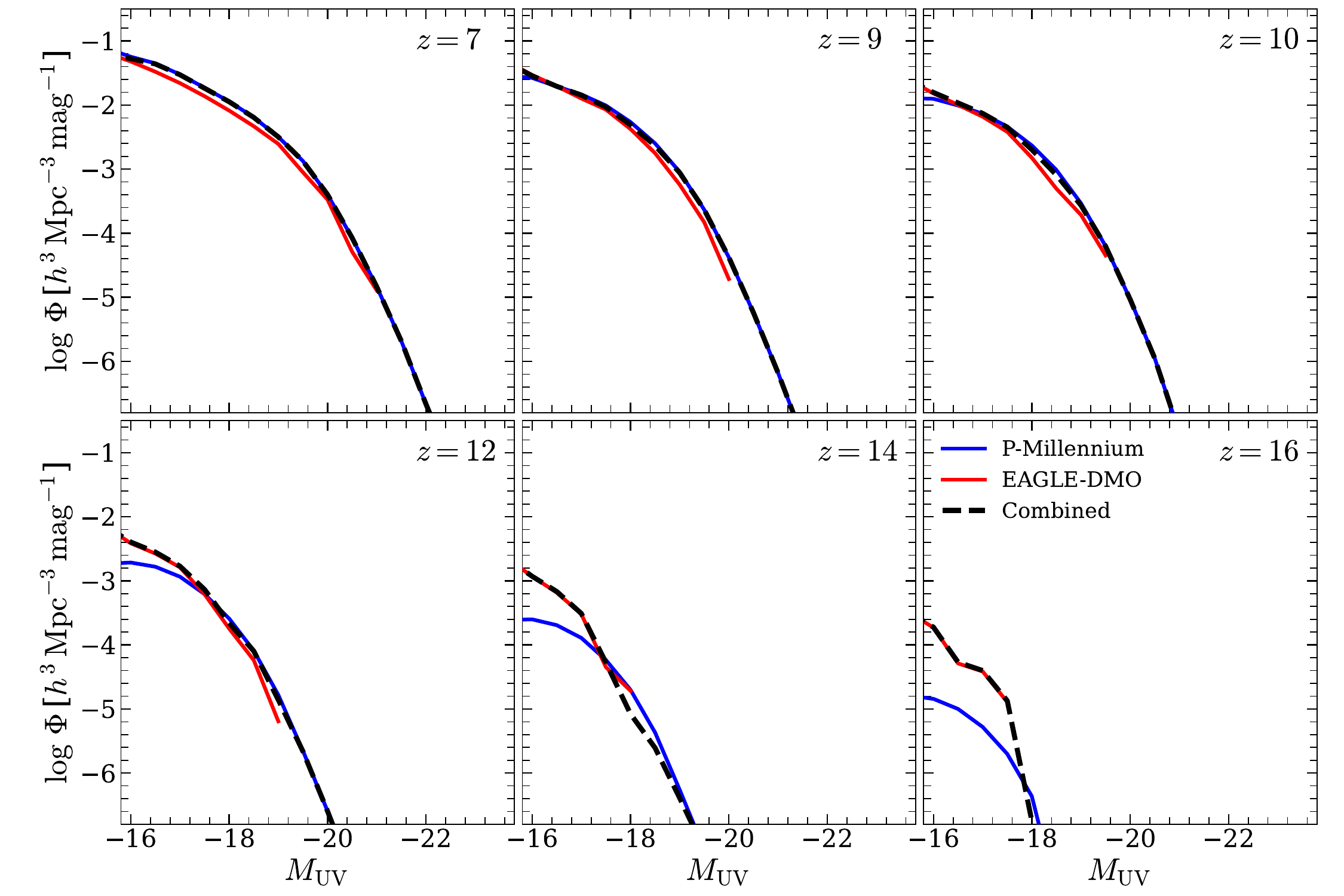}\vspace{-0.4cm}
\caption{An example of how we combine the P--Millennium and EAGLE--DMO simulations to predict the UV luminosity functions. In each panel, the result from P--Millennium is shown by the blue solid curve, that from EAGLE--DMO by the red solid curve, and the combined result by the black dashed curve.}
\label{fig:uvlf_how_to_combine}
\end{figure*}

\begin{figure*}
\centering
\includegraphics[width=\textwidth]{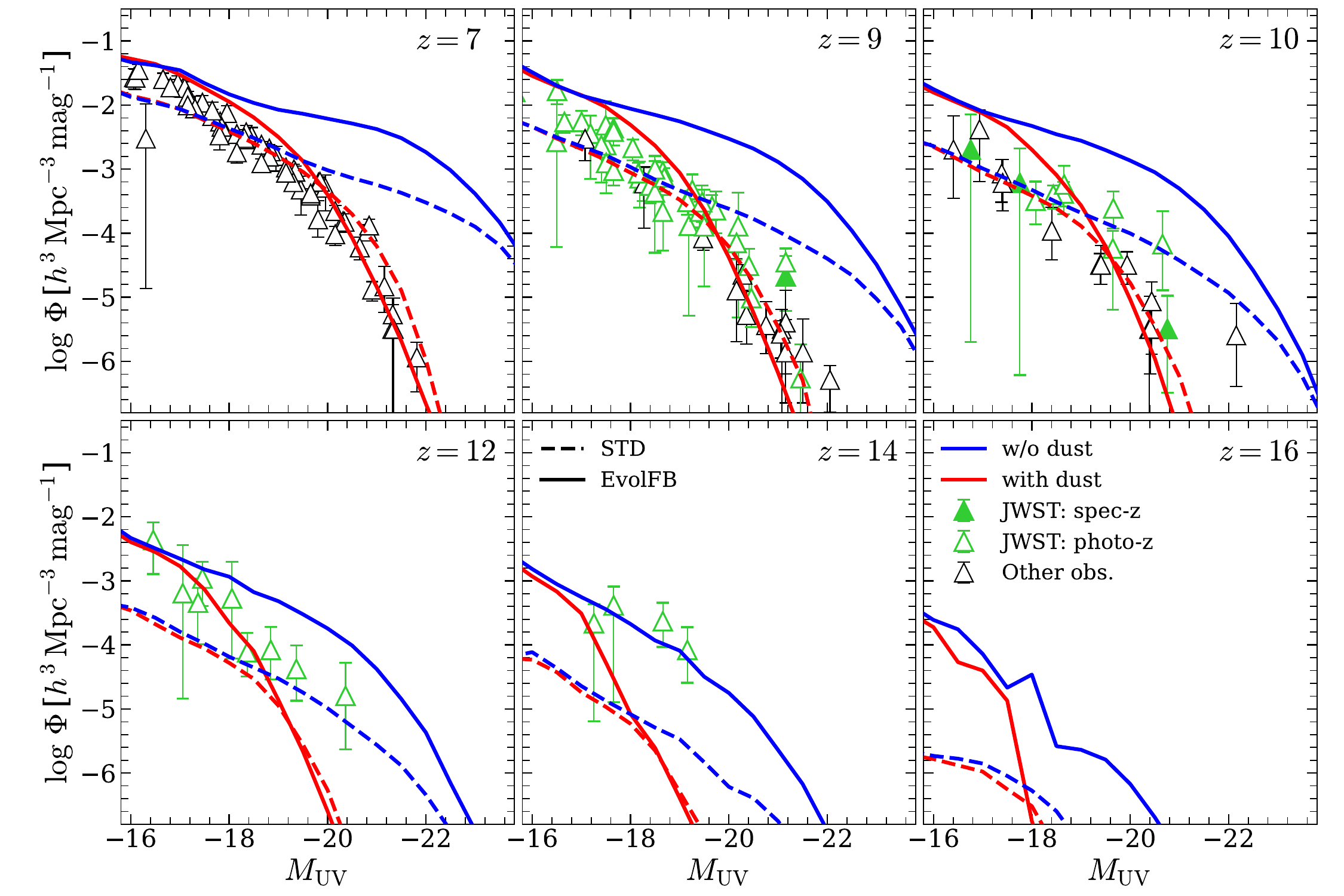}
\caption{Comparison of UV luminosity functions (UVLFs) between the predictions of \galform and observational data from $z=7$ to $z=16$ (as noted in the upper right of each panel). For the theoretical predictions, we present both the standard model (STD) and the model with evolving supernova feedback (EvolFB; \autoref{sec:model_design}), indicated by dashed and solid curves, respectively. The dust-extincted and dust-free results are shown in red and blue, respectively. In each panel, we overplot the observational data, divided into three categories: (i)~{\it JWST} data with spectroscopically confirmed redshifts (filled green triangles; \citealt{Harikane_et_al.(2024)}), (ii)~photometric redshift-based {\it JWST} data (open green triangles, including \citealt{Adams_et_al.(2023),Bouwens_et_al.(2023a),Castellano_et_al.(2023),Donnan_et_al.(2023a),Finkelstein_et_al.(2023),Harikane_et_al.(2023),Leung_et_al.(2023),Morishita_et_al.(2023),Perez-Gonzalez_et_al.(2023b),Robertson_et_al.(2023),Willott_et_al.(2023)}), and (iii)~photometric reshift-based measurements from earlier observations (open black triangles, including \citealt{Bouwens_et_al.(2011),Bouwens_et_al.(2015),Bouwens_et_al.(2021),Bowler_et_al.(2014),Bowler_et_al.(2020),Finkelstein_et_al.(2015),McLure_et_al.(2013),Morishita_et_al.(2018),Oesch_et_al.(2014),Oesch_et_al.(2018),Schenker_et_al.(2013),Stefanon_et_al.(2019)}). We exclude observational upper limits as these can be misleading.}
\label{fig:uvlf_std}
\end{figure*}

As mentioned in \autoref{sec:dmo}, in the following sections we combine the P--Millennium and EAGLE--DMO simulations to estimate the UVLFs. This combination allows us not only to study rare bright galaxies at high redshift (thanks to the large volume of P--Millennium) but also to examine the abundance of faint galaxies (thanks to the high mass resolution of EAGLE--DMO). In \autoref{fig:uvlf_how_to_combine}, we show an example of the effect of combining the two simulations at different redshifts (for simplicity, only results based on the EvolFB model are shown). P--Millennium and EAGLE--DMO are in good agreement at $z \lesssim 10$, and so the combination does not significantly affect the UVLFs at these redshifts, except for the fact that EAGLE--DMO suffers from incompleteness at the brightest end. At higher redshifts, P--Millennium shows a lower abundance at the faint end than EAGLE--DMO due to its lower mass resolution. The combined results agree well with EAGLE--DMO at the faint end and P--Millennium at the bright end. We note that, due to the very limited sample of EAGLE--DMO galaxies at $z \gtrsim 14$, the UVLFs at these redshifts exhibit noticeable fluctuations.

\autoref{fig:uvlf_std} presents the dust-extincted (red curves) UVLFs (\autoref{sec:dust_model}) predicted at $z=7-16$ by \galform based on the STD model (dashed curves) and the EvolFB model (solid curves). These are compared with observational data at the corresponding redshifts, including the new {\it JWST} observations, as well as earlier estimates using other telescopes such as the {\it Hubble Space Telescope} (HST). We use different symbols to denote different observational datasets: (i)~UVLFs from spectroscopically confirmed ( $z_{\rm spec}$) high-$z$ candidates from {\it JWST} are shown using filled green triangles \citep{Harikane_et_al.(2024)}, (ii)~UVLFs derived using photometric redshifts 
$(z_{\rm photo})$ from {\it JWST} observations are shown using open green triangles \citep{Adams_et_al.(2023),Bouwens_et_al.(2023a),Castellano_et_al.(2023),Donnan_et_al.(2023a),Finkelstein_et_al.(2023),Harikane_et_al.(2023),Leung_et_al.(2023),Morishita_et_al.(2023),Perez-Gonzalez_et_al.(2023b),Robertson_et_al.(2023),Willott_et_al.(2023)}, while (iii)~UVLFs constructed from earlier, pre-\textit{JWST} $z_{\rm photo}$ data are indicated by open black triangles \citep{Bouwens_et_al.(2011),Bouwens_et_al.(2015),Bouwens_et_al.(2021),Bowler_et_al.(2014),Bowler_et_al.(2020),Finkelstein_et_al.(2015),McLure_et_al.(2013),Morishita_et_al.(2018),Oesch_et_al.(2014),Oesch_et_al.(2018),Schenker_et_al.(2013),Stefanon_et_al.(2019)}. 

We show the comparison between the \galform predictions and the observational data up to $z=14$, and show only model predictions at $z\sim 16$, as currently there are no reliable UVLF measurements at $z>14$. \citet{Harikane_et_al.(2023)} estimated the UVLFs of galaxies at $z=16$ based on two $z_{\rm photo}\sim 16$ candidates (CR2-z16-1 identified by \citealt{Donnan_et_al.(2023a)} and S5-z16-1 identified by \citealt{Harikane_et_al.(2023)}). However, CR2-z16-1 has been spectroscopically confirmed to be at $z=4.91$ \citep{Harikane_et_al.(2024)} and S5-z16-1 may also be a galaxy located at $z=4.61$ \citep{Harikane_et_al.(2023)}. In order to make more accurate comparisons with the data, we calculated the effects of convolving the {\sc galform} predictions with a measurement error of 0.3 dex, corresponding roughly to the measurement uncertainty of observed UV magnitudes (e.g. $0.1-0.4$ dex in \citealt{Finkelstein_et_al.(2023)}). However, we find that this has a negligible effect on the predicted UVLFs. Therefore, in what follows, we simply plot the intrinsic (i.e. unconvolved with the magnitude measurement error) UVLFs predicted by \textsc{galform}.

As we can see from \autoref{fig:uvlf_std}, the EvolFB model (solid curve) always predicts a higher abundance of galaxies at the faint end across the whole redshift range ($z=7-16$). This is as expected because the evolving supernova feedback model gives weaker feedback at low masses as well as at higher redshifts \citep{Hou_et_al.(2016)}. Below $z\sim 12$, the EvolFB model predicts slightly fewer bright galaxies than the STD model. We note that the discrepancy between the STD and EvolFB models at the bright end at $z\lesssim 12$ is slightly larger than in \citet{Cowley_18}. This is because of the difference in the dust attenuation models used in this work and in \citet{Cowley_18}. Our main conclusions are unaffected by this difference. We note that the effects of dust extinction on the bright end of the UVLF are predicted to be large even at the high redshifts shown here, but we return to this issue in \autoref{sec:results-dust}.

At $z=9-10$, where there are both {\it JWST} and earlier observations (e.g. from HST), we find good agreement between the new and old measurements, as well as between $z_{\rm photo}$-based and $z_{\rm spec}$-based measurements, in agreement with the conclusions of \citet{Harikane_et_al.(2024)}. Below $z\sim 10$, the model predictions agree well with the observational data. Indeed, the STD and EvolFB models roughly bracket the data points; this is consistent with the results of \citet[][fig.~4]{Cowley_18}, even though new observational data from {\it JWST} have been added to the plot. At this point, it is important to stress again that \galform is calibrated with only low redshift constraints (see \citealt{Lacey_et_al.(2016)} and \autoref{sec:model_design} for more details) and the predictions of \cite{Cowley_18} were made prior to the availability of {\it JWST} data. Thus, the good agreement between the models and the latest UVLF measurements at these redshifts is a testament to the predictive power of the \galform model.

At higher redshifts, however, a discrepancy between the models (with dust extinction; red curves) and the observational data becomes apparent. At $z=12$, the STD model (red dashed curve) underpredicts the abundance of galaxies over the entire $M_{\rm UV}$ range, while the EvolFB model (red solid curve) agrees with the data at the faint end ($M_{\rm UV}\gtrsim -18.5$) but underpredicts the galaxy abundance at the bright end ($M_{\rm UV}\lesssim -19$) by $1-2$ dex. At $z=14$, the predictions of the EvolFB model are within the uncertainty of the data at $M_{\rm UV}> -18$, but are significantly lower at the brighter end ($M_{\rm UV}<-18$) by $2-3$ dex. These results agree with the findings of \citet{Kannan_et_al.(2023)} who compared UVLFs from {\it JWST} with a suite of simulations based on the IllustrisTNG galaxy formation model. Our results indicate that a revision of the original \galform model is required to match the observational data. This is the focus of the next section.

\section{Updating {\sc galform}}
\label{sec:update_model}
As we have just seen, the predictions made following \citet{Cowley_18}, based on two different variations of the \galform semi-analytic model within the $\Lambda$CDM framework (the STD and EvolFB models; \autoref{sec:method}), are in good agreement with the latest {\it JWST} UVLF data for $z\lesssim 10$. However, the models underpredict the abundance of galaxies at the bright end of the UVLF ($M_{\rm UV}<-18$) at higher redshifts ($z\sim 12-14$) when dust extinction is included. This discrepancy could be due to several factors. Here, we explore the sensitivity of the predicted UVLFs to the details of specific physical components of the model -- most notably, the dust model and the IMF (described in \autoref{sec:method}). We investigate whether reasonable adjustments of the model parameters make it possible to bring the \galform predictions into closer agreement with the observed UVLFs at $z\gtrsim12$. 

\subsection{The effect of dust attenuation}
\label{sec:results-dust}
One way to resolve the difference between our model predictions and the observed UVLFs at high redshifts is to alter the importance of dust attenuation, since dust in galaxies absorbs light and thus reduces the number counts of bright galaxies. As a limiting case, we begin by turning off the dust attenuation in \galform altogether (for both the STD and EvolFB models) and calculate dust-free UVLFs at $z=7-16$; these predictions are also shown in \autoref{fig:uvlf_std} (blue curves).

We can see that while switching to the dust-free models (blue curves) does not change the galaxy abundance at $M_{\rm UV}\sim -16$, there is an impact at the bright end, where the abundance is significantly enhanced. This is as expected since, at fixed redshift, bright galaxies are typically more massive and have experienced more star formation than faint galaxies, resulting in more metals and, consequently, more dust production (e.g. \citealt{Dayal_et_al.(2022),Sommovigo_et_al.(2022)}). At $z\lesssim 10$, where the original models previously agreed well with observations, the dust-free models now significantly overestimate the abundance of galaxies. At $z=12$, the dust-free STD model (blue dashed curve) is in agreement with the data at the bright end ($M_{\rm UV}\lesssim -18$), but still underestimates the abundance of galaxies at the faint end. The dust-free EvolFB model (blue solid curve) agrees well with the data at the faint end ($M_{\rm UV}\gtrsim -18$), but overestimates the galaxy abundance at the bright end. At $z=14$, only the dust-free EvolFB model (blue solid curve) agrees with the data. Our results suggest that an extra process that causes the importance of dust attenuation to vary with redshift may be one way to explain the discrepancy between models and observations. This process should preserve the amount of dust attenuation of the original model at $z\lesssim 10$, reduce it at $z=12$ and make it largely negligible at $z\sim 14$.

There is some observational evidence that dust extinction may be less important at very high redshift than in the local Universe (e.g. \citealt{Austin_et_al.(2024),Cullen_et_al.(2024),Morales_et_al.(2024),Topping_et_al.(2024)}). Several studies \citep[e.g.][]{Vladilo_et_al.(2011),De_Cia_et_al.(2013),De_Cia_et_al.(2016),Wiseman_et_al.(2017)} have reported a metallicity-dependent dust-to-metal ratio (DTM), which is about 50\% lower at metallicity $Z=0.1Z_{\odot}$ than at solar metallicity, $Z_{\odot}$ (see \citealt{Maiolino_et_al.(2019)} for a review). Since metallicities are lower at high redshift, this would be expected to lead to a lower DTM. Contrary to this, \citet{Wiseman_et_al.(2017)} reported that they did not find any obvious redshift-dependence of the DTM; however, this conclusion is based on small galaxy samples up to a redshift of only $z\sim 5$. In {\sc galform}, a constant DTM is assumed with a value of 0.334 \citep{Lacey_et_al.(2016)} in order to match the dust-to-gas ratio in the solar neighbourhood, $6.7\times 10^{-3}$ for $Z_{\odot}=0.02$ \citep{Silva_et_al.(1998)}. If the reports of a lower DTM ratio at lower metallicity are correct, \galform would overestimate the dust content at low metallicity and, consequently, that of galaxies at high redshifts. 

\begin{figure*}
\centering
\includegraphics[width=\textwidth]{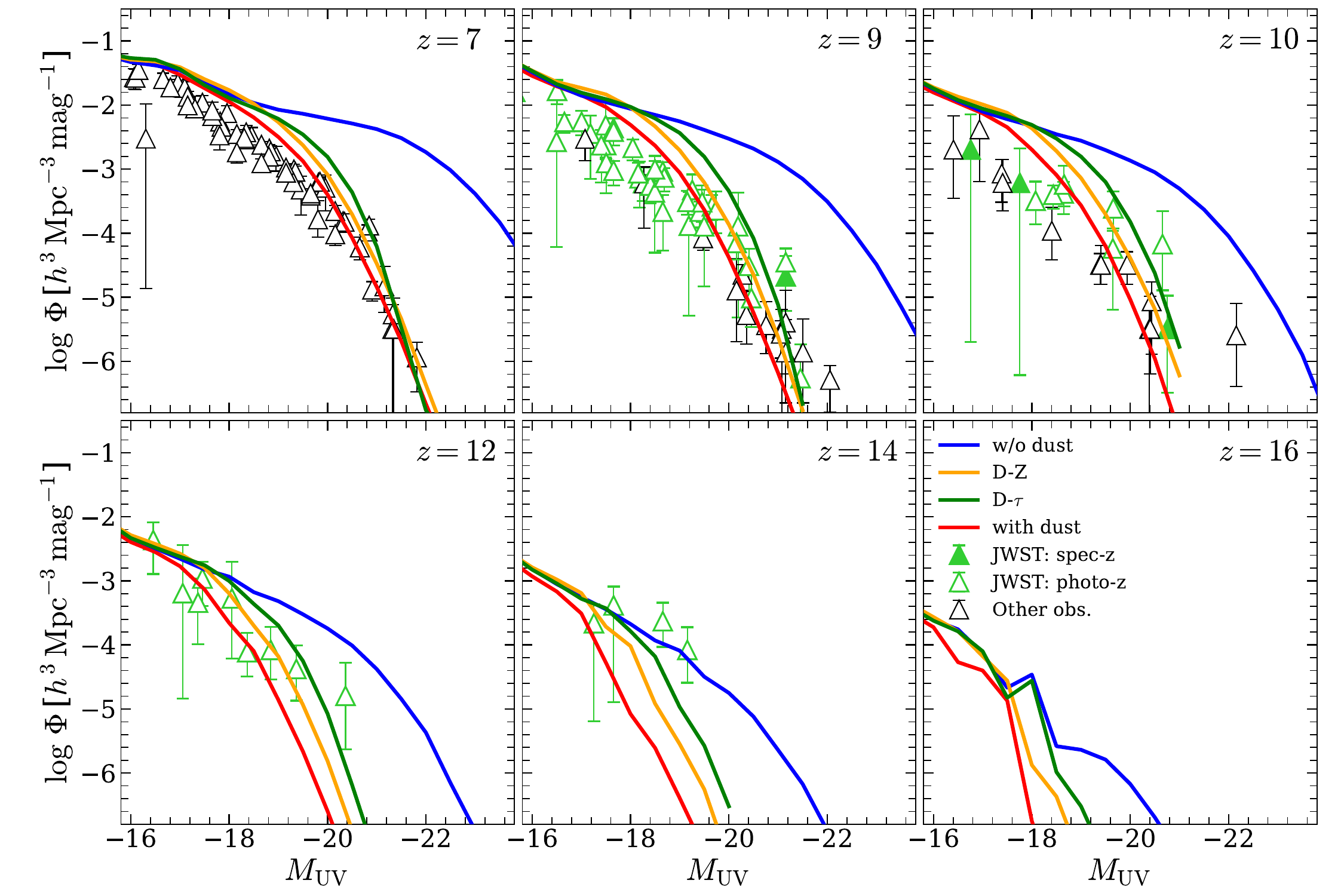} \vspace{-0.4cm}
\caption{The effect of different assumptions for the dust production on the predicted UVLFs (based on the EvolFB model; \autoref{sec:sn_model}). In each panel, the results for (i) the original dust-extincted model (\autoref{sec:dust_model}), (ii) the dust-free model, (iii) the D-$\tau$ assumption (i.e., dust grains grow gradually, on a timescale, $\tau_{\rm growth}$), and (iv) the D-Z assumption (i.e., the dust-to-metal ratio varies with metallicity) are indicated by red, blue, green, and orange curves, respectively. The other symbols are as in \autoref{fig:uvlf_std}.}
\label{fig:uvlf_dust}
\end{figure*}

Another, more radical, possibility is that dust may not have had time to form, in some of the high redshift galaxies (e.g. \citealt{Markov_et_al.(2024)}). The details of dust formation and destruction mechanisms are very uncertain even at low redshift but, from analytical and observational considerations, \citet{Mattsson_et_al.(2012)} concluded that dust growth in the interstellar medium is the most important dust formation mechanism. They proposed a dust grain growth timescale, $\tau_{\rm growth}$, which is found to increase with decreasing metallicity and can be as long as $\gtrsim 1\,\mathrm{Gyr}$ for the lowest metallicity galaxies \citep{Galliano_et_al.(2021)}. Since $z = 14$ corresponds to only $\sim 300\,\mathrm{Myrs}$ after the Big Bang, galaxies at this redshift may not have had enough time to make a significant amount of dust. \citet{Schneider_et_al.(2024)} proposed a parametrization of $\tau_{\rm growth}$ which depends on the distribution of grain sizes, and on the density, temperature, and metallicity of the interstellar medium. This model also shows a negative correlation with metallicity, similar to that found by \citet{Galliano_et_al.(2021)}.

We extend the original dust model in {\sc galform} (\autoref{sec:dust_model}), including these effects through two alternative assumptions: 

(i) D-$\tau$ assumption: dust forms gradually from metals in the cold interstellar medium on a timescale, $\tau_{\rm growth}$, that depends on certain properties of the gas and dust (e.g. temperature, metallicity, dust grain size). We adopt the parametrization suggested by \citet[][their eq.~23]{Schneider_et_al.(2024)} and fix all the parameters, except the gas metallicity, to their fiducial values:
\begin{equation}
\label{eq:tgrowth}
\tau_{\rm growth} = 6.7\,\mathrm{Myr}\times \left(\frac{Z}{Z_{\odot}}\right)^{-1},
\end{equation}
where $Z$ is the gas-phase metallicity of the galaxy. We assume that the DTM increases at a constant rate and reaches the value, $\mathrm{DTM}_{\odot}$, as assumed in the original {\sc galform} dust model (\autoref{sec:dust_model}) after a time $\tau_{\rm growth}$. Thus, for a given galaxy, only a fraction, $\tau_{\rm form}/\tau_{\rm growth}$ ($\leqslant 1$), of the dust is able to form, where $\tau_{\rm form}$ is the characteristic time during which dust has been able to form. We take this characteristic time to be the gas depletion time of the galaxy, $\tau_{\rm dep}$, given by:
\begin{equation}
\label{eq:depeltion_time}
\tau_{\rm dep} = \frac{M_{\rm cold}}{(1-R+\beta)\psi},
\end{equation}
where $M_{\rm cold}$ is the cold gas mass of the galaxy, $\psi$ is the star formation rate (SFR), $\beta \psi$ is the rate at which gas is ejected from the cold component into the halo reservoir by SN feedback, $\beta$ is the mass loading factor given by \autoref{eq:sn_updated_beta}, and $R\psi$ is the rate at which gas is returned into the cold component by stellar evolution; the returned fraction, $R$, is determined by the IMF ($R=0.44$ for the \citealt{Kennicutt(1983)} IMF and $R=0.54$ for the tilted, $x=1$, IMF). As \galform adopts different IMFs for different SF modes (\autoref{sec:imf_model}), we calculate the fraction of dust formed ($\tau_{\rm dep}/\tau_{\rm growth}$) for the quiescent and starburst SF modes separately. At high redshift, galaxies have longer dust grain growth timescales because of their lower metallicity, resulting in less attenuation than in the original {\sc galform} dust model. 

(ii) D-Z assumption: the dust-to-metal ratio varies with the gas-phase metallicity. We assume that the DTM (normalized by the MW value) follows a simple relation with the gas-phase metallicity as advocated by \citet{Wiseman_et_al.(2017)}:
\begin{equation}
\label{eq:dtm}
\frac{\mathrm{DTM}_{\hphantom{\odot}}}{\mathrm{DTM}_{\odot}} = 0.4\times[Z/H]+1, 
\end{equation}
where $[Z/H] = \log(Z/Z_{\odot})$. At $[Z/H]<-2.5$ and $[Z/H]>0$, the normalized DTM is set to zero (i.e. no dust) and one (i.e. the dust-to-metal ratio has the same value as in the MW) respectively. Thus, at higher redshift, where the gas is, on average, more metal-poor than at lower redshift, the DTM and, in turn, the dust mass is lower and dust attenuation is weaker than in the original model.

In \autoref{fig:uvlf_dust} we show predictions for the different assumptions about the dependence of dust mass on metallicity for the EvolFB model (\autoref{sec:sn_model}), alongside the original dust-extincted (red curve) and the dust-free models (blue curve). The two assumptions, one, that the DTM decreases with decreasing metallicity (the D-Z assumption), and two, that at lower gas metallicity dust takes longer to form (the D-$\tau$ assumption) increase the observed galaxy luminosities and bring the model into better agreement with the data at all redshifts. In both alternatives, the amount of DTM builds up with time and approaches the original model (in which the DTM is independent of redshift and metallicity) at low redshift. At $z=12$, the D-$\tau$ model (green) gives an excellent match to the data, while the D-Z case (orange) only slightly underestimates the abundance at $M_{\rm UV}\sim -20.4$. At $z=14$, the D-$\tau$ model slightly underpredicts the counts of the brightest galaxies. 

Our two variable DTM models are simplified but have no free parameters. It could even be that the physical origin of both is the same, since 
the metallicity dependence of the DTM could be due to the fact that dust needs more time to form at low metallicity. These models serve to illustrate how plausible assumptions regarding the dependence of dust grain formation on metallicity -- for which there is both theoretical and observational support -- can bring our model predictions into agreement with the observations even at the highest redshifts: there is no conflict between the {\it JWST} discovery of galaxies at $z\gtrsim 12$ and the standard $\Lambda$CDM model.

\subsection{The effect of the initial stellar mass function}
\label{sec:results-imf}
\begin{figure*}
\centering
\includegraphics[width=\textwidth]{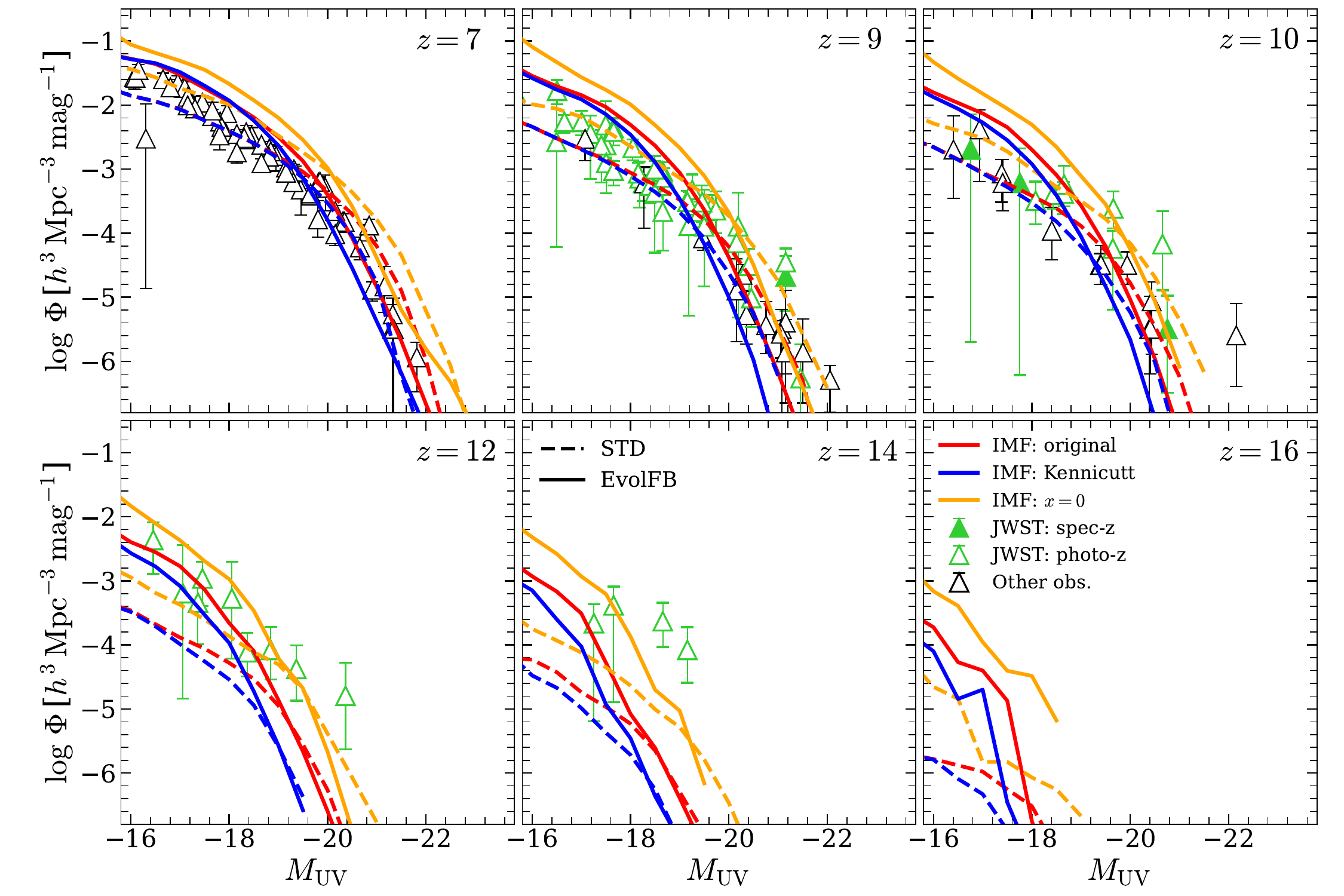}\vspace{-0.6cm}
\caption{The effect of varying the stellar initial mass function (IMF) on the predicted UVLFs. In each panel, models with (i) the original IMF set-up (a Kennicutt IMF in discs and a top-heavy, $x=1$, IMF in starbursts; see \autoref{sec:imf_model}), (ii) a universal Kennicutt IMF for all SF, and (iii) an extremely top-heavy universal IMF ($x=0$) also for all SF, are indicated by red, blue and orange curves, respectively (dashed curves for the STD model and solid for the EvolFB model). The other symbols are as in \autoref{fig:uvlf_std}.}
\label{fig:uvlf_imf}
\end{figure*}

Another possible mechanism to boost the UV luminosity of galaxies at high redshifts is to adopt a top-heavy IMF at these redshifts: massive stars emit more UV light than low-mass stars for a given amount of stellar mass formed. As mentioned in \autoref{sec:imf_model} (see also \citealt{Baugh_05, Lacey_et_al.(2016)}), \galform adopts two different IMFs in different star formation (SF) modes: a \citet{Kennicutt(1983)} IMF for quiescent (disc) SF and a top-heavy IMF ($x=1$; see \autoref{eq:imf}) in starbursts. This set-up is kept unchanged throughout the entire evolution. To test the impact of the IMF on the predicted UVLFs, we re-ran \galform with two more IMF variations: (i)~a \citet{Kennicutt(1983)} IMF in both SF modes (quiescent and starburst) and (ii)~an extremely top-heavy IMF ($x=0$; see \autoref{eq:imf}) in both SF modes.

In \autoref{fig:uvlf_imf}, we present the UVLFs from $z=7$ to $z=16$ for all three IMF variations. In all cases, the UVLFs include attenuation by dust as in the fiducial version of {\sc galform}. Note that the production of metals is calculated self-consistently for the chosen IMF, and is larger for more top-heavy IMFs (see \citealt{Lacey_et_al.(2016)} for more details), and so the amount of dust also depends on the IMF. In both \galform models (STD and EvolFB), the extremely top-heavy IMF case (orange curves) predicts the highest abundance of galaxies and the case of a \citet{Kennicutt(1983)} IMF in both SF modes (blue curves) predicts the lowest abundance. This shows that the dual mode IMF implementation in the original \galform model (where the redshift dependence arises implicitly because starbursts are more common at high redshift) is already an improvement over the assumption of a single \citet{Kennicutt(1983)} IMF in all SF modes. Clearly, however, the effect is not large enough to account fully for the discrepancy between the model and the observed UVLFs. 

At $z=12$, where the original EvolFB model (red solid curve) agrees with observations at the faint end, the discrepancy is reduced further with the extremely top-heavy IMF (orange solid curve); however, the predicted abundance at $M_{\rm UV}\sim -20.4$ is still lower than the observational data point by an order of magnitude. At $z=14$, the discrepancy between the predicted and observed UVLFs is also smaller when adopting the extremely top-heavy IMF, but the difference is still significant. In conclusion, although a top-heavy IMF at high redshifts ($z\sim 12-14$) helps to ease the tension between models and {\it JWST} data, this alone cannot account for the whole discrepancy.

\subsection{Comparisons with other studies}
\label{sec:results-compare}

\begin{figure*}
\centering
\includegraphics[width=0.9\textwidth,trim={0.5cm 0.4cm 0 0}, clip]{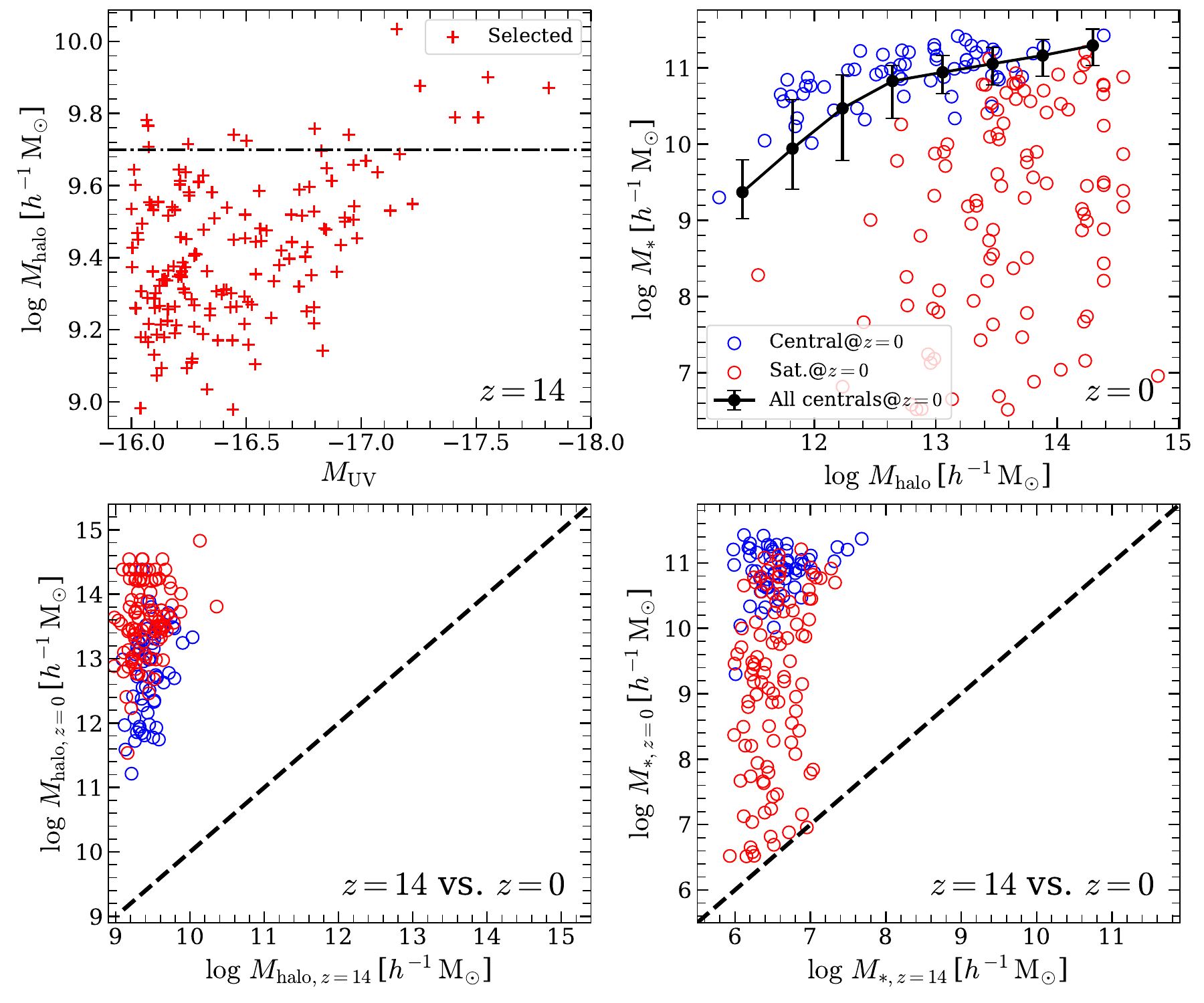} \vspace{-0.1cm}
\caption{Predicted properties of high-$z$ ($z=14$) bright galaxies and their descendants at $z=0$. Upper left: the distribution of bright galaxies ($M_{\mathrm{UV}} \leqslant -16$, where $M_{\mathrm{UV}}\equiv M_{\rm AB}(1500\Angstrom)-5\log h$ is the absolute UV magnitude of the galaxies at $z=14$) in the ($M_{\rm UV}$,$M_{\rm halo}$) plane at $z=14$. The red plus symbols indicate the distribution of randomly selected samples. The black dashed-dotted line indicates the threshold, $M_{\rm halo}=5\times 10^{9}\,h^{-1}\,\mathrm{M_{\odot}}$, above which we take halos from the P--Millennium simulation and below which we take halos from EAGLE-DMO (see \autoref{sec:dmo} for more details). Upper right: the correlation between halo mass and stellar mass of the $z=0$ descendants of the randomly selected galaxies. The blue circles represent descendants that are centrals at $z=0$ and red ones show satellites at $z=0$. The black curve indicates the halo mass-stellar mass relation for all centrals at $z=0$ (for $\log\,(M_{\rm halo}/h^{-1}\,\mathrm{M_{\odot}})>11.2$). Lower left: the correlation between the halo mass of the selected galaxies at $z=14$ and their $z=0$ descendants with blue circles representing galaxies which end up as centrals and red ones as satellites. Lower right: the correlation between the stellar mass of the selected galaxies at $z=14$ and their descendants at $z=0$ with symbols as in previous panels. The black dashed lines in the bottom two panels indicate $y=x$. Nearly all randomly selected high-$z$ bright galaxies were centrals at $z=14$.}
\label{fig:prediction_1}
\end{figure*}

\begin{figure}
\centering
\includegraphics[width=\columnwidth,trim={0, 0.5cm, 0, 0}, clip]{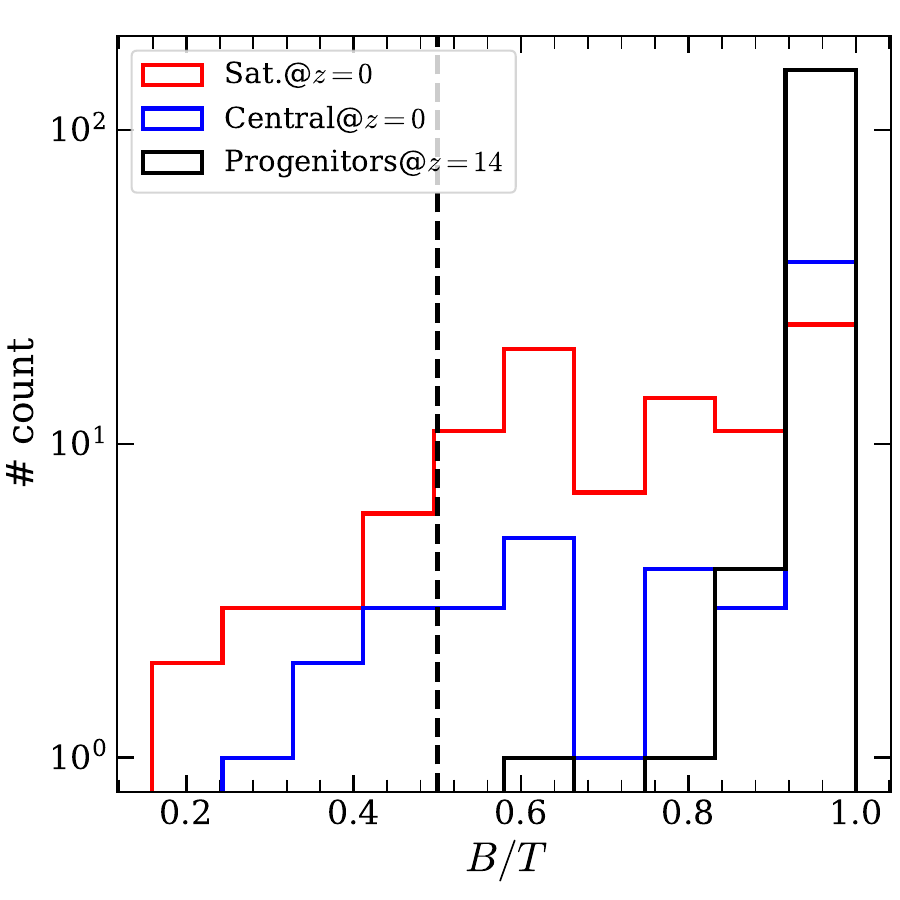} \vspace{-0.4cm}
\caption{Distributions of the bulge-to-total stellar mass ratio ($B/T$) of (i)~the randomly selected sample at $z=14$ (black), (ii)~ the descendants at $z=0$ of the selected samples which end up as central galaxies (blue), and (iii)~the descendants at $z=0$ of the selected samples which end up as satellites (red). The black dashed line indicates $B/T=0.5$.}
\label{fig:prediction_2}
\end{figure}

In the previous two sections, we investigated the influence of dust attenuation and the stellar IMF in predictions of the UVLFs using our semi-analytic model, {\sc galform}. We find that a model in which dust attenuation is insignificant at $z\gtrsim 12$ and in which supernova feedback is relatively weaker at high redshift (the EvolFB model; see \citealt{Hou_et_al.(2016)}) is able to remove the discrepancy between observations and the standard \galform model up to the highest redshift for which there is reliable data, $z=14$. An extremely top-heavy IMF at such high redshifts can reduce the discrepancy even if dust is present, although it alone cannot entirely resolve the differences.

Previous studies have also considered these two factors as potential solutions to the puzzle of the high-redshift {\it JWST} galaxies. For example, \citet{Ferrara_et_al.(2023)} claimed, using an empirical model, that they were able to reproduce the UVLFs up to $z\sim 14$ if galaxies at $z\gtrsim 11$ contain negligible amounts of dust, which is roughly consistent with our findings. \citet{Iocco_et_al.(2024)}, also using empirical models, suggested a negligible role of dust extinction at the highest redshifts, but stated that a modification of the star formation rate to incorporate a larger fraction of luminous objects per massive halo is also required to match the observations. However, \citet{Mauerhofer_et_al.(2023)}, using a simple semi-analytic model tuned to match data at $z = 5-9$, claimed that dust does not have a significant impact on the visibility of early galaxies at $z \gtrsim 12$ and, in fact, reported an increase in dust content with increasing redshift at a given stellar mass. Their model underpredicts the observed UVLF at $z \sim 12-18$ by a large factor. Using a simple empirical model, \citet{Wang_et_al.(2023)} reported that bright UVLFs at $z\sim 13$ can be fitted by a model with no dust attenuation but argued that a contribution to the luminosity from Population III stars is necessary (see also \citealt{Ventura_et_al.(2024)}). 

\citet{Yung_et_al.(2024)} studied the effect of the IMF on the UVLFs using the Santa Cruz semi-analytic model by simply scaling up the predicted UVLFs, which are based on a \citet{Chabrier2023}-like IMF, by a factor of several. They claimed that a top-heavy IMF with a boost of a factor of $\sim 4$ to the UV luminosities can bring the  predictions into agreement with the observations. We note that the simple scaling that they adopted does not properly capture all the effects of the IMF, such as the change in dust attenuation with different IMFs. We choose to implement a self-consistent IMF change in {\sc galform}. \citet{Trinca_et_al.(2024)} also found that a metallicity and redshift-dependent IMF can reduce the discrepancy between models and observations. \citet{Rasmussen_Cueto_et_al.(2023)}, however, argued that a top-heavy IMF alone could not solve the problem of the high abundance of bright galaxies at high redshift. They claimed that although a top-heavy IMF produces more UV photons for a given amount of stellar mass, it will also give rise to stronger SN feedback, which suppresses star formation and hence keeps the UVLFs nearly unchanged compared to the results with a standard \citep{Salpeter(1955)} IMF.

The studies above have two common themes: dust attenuation and a top-heavy IMF. Other explanations for the existence of bright galaxies at high redshift have been proposed. For example, \citet{Dekel_et_al.(2023)} and \citet{Li_et_al.(2023)} claimed that an enhanced star formation efficiency (SFE) from feedback-free starbursts (FFS) at high redshift may enhance the abundance of massive galaxies (see also \citealt{Qin_et_al.(2023)}). They do not, however, address the critical point of what the present-day luminosity function of galaxies would look like in such a scenario.
\citet{Sun_et_al.(2023)} and \citet{Shen_et_al.(2023)} claimed that scatter in UV luminosity at a fixed halo mass alone may be enough to resolve any disagreements at high redshift (see also \citealt{Yung_et_al.(2024)}), although they do not explain the physical source of such scatter. Furthermore, \citet{Mirocha_et_al.(2023)} warned that the upscattering of low-mass halos into bins of brighter galaxies will introduce additional tensions because the resulting stellar ages, masses, and spectral slopes would be much lower than indicated by observations. To solve these extra tensions requires both star formation and dust production to be more efficient than expected at $z\gtrsim 10$, which conflicts with the expectations of previous studies (e.g. \citealt{Ferrara_et_al.(2023),Iocco_et_al.(2024),Wang_et_al.(2023)}), as well as this work. Note that in models like {\sc galform}, scatter in the UV luminosity at a given halo mass arises automatically from the physics of galaxy formation, since halos of the same mass have different assembly histories, leading to different formation histories for the galaxies they contain.

The above studies focused exclusively on the galaxy population at high redshift and either employed empirical models (e.g. \citealt{Ferrara_et_al.(2023),Wang_et_al.(2023),Iocco_et_al.(2024)}) or adopted simple scalings in semi-analytic models (e.g. \citealt{Yung_et_al.(2024)}). By contrast, here, we employ an {\em ab initio} model of galaxy formation based on physical principles rather than empirical prescriptions, which incorporates the effects of a variable IMF, dust etc. in a self-consistent way. Most importantly, our model follows galaxy formation and evolution throughout the whole of cosmic history and accounts for a wide range of observational data at all times, not just at high redshift.

\section{The descendants at $z=0$ of the high-redshift {\it JWST} galaxies}
\label{sec:descendants}

In \autoref{fig:uvlf_dust}, we saw that the EvolFB model with a decreasing dust content towards high redshift approximately match the high galaxy abundance measured by {\it JWST} at $z=12-14$. In this section, we trace the bright galaxies at $z=14$ in the EvolFB model to $z=0$ and study the properties of the descendants of these high-$z$ bright galaxies. The results for the STD model are similar.

To build up the sample, we first consider all the EAGLE--DMO galaxies with $M_{\rm UV} < -16$ and $M_{\rm halo} < 5 \times 10^9\,h^{-1}\,\mathrm{M_{\odot}}$ at $z = 14$. Due to the larger volume of P--Millennium, we randomly select $1/512$ (the volume ratio of the two simulations) of the galaxies with $M_{\rm UV} < -16$ and $M_{\rm halo} > 5 \times 10^9\,h^{-1}\,\mathrm{M_{\odot}}$ at $z = 14$ from P--Millennium to match the sample from EAGLE--DMO. Thus, we have 161 galaxies in total, covering a wide halo mass range. In the upper left panel of \autoref{fig:prediction_1}, we show the distributions of the selected galaxies (red plus symbols) in the ($M_{\rm UV}$, $\log\,M_{\rm halo}$) plane at $z=14$. Apart from the expected increasing trend of $M_{\rm halo}$ towards the UV-brighter end, we see that galaxies with larger halo masses exhibit a greater scatter in UV magnitude. 

The other panels of \autoref{fig:prediction_1} show (i)~the relation between stellar mass and halo mass of the $z=0$ descendants of the selected $z=14$ bright galaxies (upper right), (ii)~the relation between the halo mass of the selected $z=14$ bright galaxies and their descendants at $z=0$ (lower left), and (iii)~the relation between the stellar mass of the selected $z=14$ bright galaxies and their $z=0$ descendants (lower right). The $z=14$ galaxies are seen to typically reside in massive halos at $z=0$, with the halo mass of the descendants spanning a wide range, from $10^{11}\,h^{-1}\,\mathrm{M_{\odot}}$ to $10^{15}\,h^{-1}\,\mathrm{M_{\odot}}$ with a median of $\sim 2.5 \times 10^{13}\,h^{-1}\,\mathrm{M_{\odot}}$, consistent with \citet{Chen_et_al.(2023)}. Interestingly, the stellar mass of the descendants spans an even wider range, from dwarfs of $3.2\times 10^{6}\,h^{-1}\,\mathrm{M_{\odot}}$ to very massive galaxies of $3.2 \times 10^{11}\,h^{-1}\,\mathrm{M_{\odot}}$. 

We find that although nearly all selected galaxies were centrals at $z=14$ (159 out of 161 selected galaxies), over half (101 out of 161) become satellites (red circles) in more massive halos at $z=0$. These satellites are captured by massive halos at different redshifts, have their gas stripped, star formation suppressed, and stellar mass growth stopped, resulting in the wide span of stellar mass at $z=0$ seen in the figure. The descendants of bright high-$z$ galaxies that are centrals at $z=0$ populate the high-mass tail of the stellar-halo mass relation of all central galaxies at $z=0$.

 We present the distributions of the bulge-to-total stellar mass ratio ($B/T$) in \autoref{fig:prediction_2}  for (i)~the $z=14$ selected galaxies, (ii)~descendants at $z=0$ which end up as central galaxies at $z=0$, and (iii)~descendants at $z=0$ which are satellites at $z=0$. As can be seen, the selected $z=14$ galaxies are typically extremely bulge-dominated ($B/T\sim 1$). This reflects the fact that in the model, most of these galaxies are starbursts triggered by either disc instabilities or major galaxy mergers, and these processes are assumed to create new stars in the bulge and destroy any pre-existing stellar disc. At $z=0$, however, $B/T$ becomes on average lower than at $z=14$, indicating disc formation after $z=14$. Galaxies which end up as satellites are more likely to have more disc-dominated morphologies than those that become centrals.

\section{Conclusions and Discussion}
\label{sec:conclusion}
Numerous observational papers reporting the discovery of galaxies at very high redshift by {\it JWST} have claimed, with little justification, that the abundance and/or brightness of these galaxies cannot be accommodated within the $\Lambda$CDM model (see references in the Introduction). Here, we have compared pre-existing predictions for the UV luminosity functions of high redshift galaxies to the {\it JWST} data. These predictions, based on the Durham semi-analytic model of galaxy formation, {\sc galform}, applied to merger trees from the P--Millennium N-body simulation, were presented in \citet{Cowley_18} five years prior to the launch of {\it JWST}. In this work, we compare {\galform} predictions with {\it JWST} observations, using a combination of the large-volume  P--Millennium N-body simulation and high-resolution EAGLE dark matter-only (DMO) simulation. The predictions generally agree well with the data, particularly when the model is extended to account for the fact that the timescale for the growth of dust grains is of the same order as the age of the Universe at the highest redshifts probed by {\it JWST}, $z\gtrsim 12$. We extended our original model by considering two different ways for the dust-to-metal ratio, DTM, depends on metallicity. Our main conclusions are:
\begin{enumerate}
\item The original \galform model of \citet{Cowley_18} successfully predicted the observed UV luminosity functions out to $z\sim 10$, including the new data from {\it JWST}. 

\item At higher redshifts, $z\sim 12-14$, the original model underpredicts the abundance of galaxies, especially at the bright end (\autoref{fig:uvlf_std}). If dust attenuation is ignored, however, the model matches the data at $z=14$ and overpredicts it at $z=12$. Including either of the two extensions that consider the dust growth timescale (particularly, the D-$\tau$ model) results in a close match to the UVLFs at all redshifts. 

\item The extended {\sc galform} model predicts the existence of a sizeable population of bright galaxies at $z=16$, for example, an abundance of $\simeq 10^{-6}\,h^3\, {\rm Mpc}^{-3} {\rm mag}^{-1}$ at $M_{\rm UV}=-19$ (\autoref{fig:uvlf_dust}).

\item A top-heavy IMF is essential to match the {\it JWST} UV luminosity functions at high redshift (\autoref{fig:uvlf_imf}).

\item The galaxies seen by {\it JWST} at $z=14$ have stellar masses in a narrow range ($\sim 10^{6}-10^{7}\,h^{-1}\,\mathrm{M_{\odot}}$) and reside in halos with a narrow range of mass 
($\sim 10^{9}-10^{10}\,h^{-1}\,\mathrm{M_{\odot}}$). In contrast, their descendants at $z=0$ span a wide range of stellar ($\sim 3.2\times 10^{6}-3.2\times10^{11}\,h^{-1}\,\mathrm{M_{\odot}}$)
and halo ($\sim 10^{11}-3.2\times10^{14}\,h^{-1}\,\mathrm{M_{\odot}}$) masses. 
Although the {\it JWST} galaxies are all central galaxies at $z=14$, over half of them have become satellites in more massive halos by the present day (\autoref{fig:prediction_1}).
\item The $z=14$ bright galaxies are typically bulge-dominated ($B/T\sim 1$), while their descendants at $z=0$ have significant disc components, especially those that end up as satellites (\autoref{fig:prediction_2}). 
\end{enumerate}

We have shown in this paper that, contrary to many statements in the recent literature, the galaxies discovered by {\it JWST} at very high redshift are a natural expectation in the $\Lambda$CDM model. This result was already anticipated in the predictions published by \cite{Cowley_18} in advance of the observations, although to obtain a close match to the new data we had to extend the model to take into account the growth timescale of dust grains which is comparable to the age of the universe at the highest redshifts probed by {\it JWST}. 

Current models of galaxy formation, be they semi-analytic or hydrodynamics simulations, include a number of adjustable parameters that describe uncertain physical processes and that must be fixed (``calibrated", in the language of hydrodynamics simulations). From its inception over 20 years ago, the strategy for {\sc galform} has been to fix those parameters by requiring agreement with a small subset of data, predominantly at $z=0$. Thus, outputs of the model at higher redshifts are genuine theoretical predictions. Comparisons with new data may then require modifying the model. In our case, we had to extend the dust model, as described above, in order to match the observed UV luminosity functions at $z=14$. The results for even higher redshifts presented here are our new predictions that await testing against data. A remarkable outcome of this work is that we have presented a model that matches a wide variety of observed galaxy properties and scaling relations, not only at the redshifts of the new {\it JWST} data but at all redshifts, including the present day.

\section*{Acknowledgements}
We thank the anonymous referee for helpful comments that led to improvements in the paper. We thank Philip Wiseman for explaining the metallicity-dependent dust-to-metal ratio and Zefeng Li for discussions on dust growth history. We acknowledge support from STFC (ST/X001075/1). SB is supported by a UK Research and Innovation Future Leaders Fellowship (MR/V023381/1).
CSF acknowledges support from the European Research Council through Advanced Investigator grant DMIDAS (GA 786910). This work used the DiRAC@Durham facility managed by the Institute for Computational Cosmology on behalf of the STFC DiRAC HPC Facility (www.dirac.ac.uk). The equipment was funded by BEIS capital funding via STFC capital grants ST/K00042X/1, ST/P002293/1, ST/R002371/1, and ST/S002502/1, Durham University and STFC operations grant ST/R000832/1. DiRAC is part of the UK National e-Infrastructure.
 
\section*{Data availability}
The predicted ultraviolet luminosity functions from {\sc galform} were included as supplementary materials and can be obtained from the journal website and \url{https://icc.dur.ac.uk/data/}.

\bibliographystyle{mnras}
\bibliography{ref}



\label{lastpage}
\end{document}